\documentclass[reprint,twocolumn,groupedaddress,showkeys,floatfix,superscriptaddress,aps,pre]{revtex4-2}
\usepackage[colorlinks=true,citecolor=red,filecolor=green,linkcolor=blue,pdfnewwindow=true]{hyperref}
\usepackage{amsmath} 
\usepackage[version=4]{mhchem}
\usepackage{graphicx}
\usepackage{bm}
\usepackage[title]{appendix} 
\usepackage{float}
\usepackage[normalem]{ulem}
\usepackage{setspace}
\usepackage{xcolor}
\usepackage{hyperref}
\usepackage{natbib}
\usepackage{footnote}
\usepackage{soul}
\usepackage{mathtools}
\usepackage{booktabs}

\usepackage{atbegshi}
\usepackage{placeins}

\makeatother
\begin{document}
	
\title{Human brain state classification via permutation entropy of EEG phase dynamics across consciousness levels and inattentive-type ADHD}

	\author{Athokpam~Langlen~Chanu}             \email{athokpam.chanu@apctp.org}
	\affiliation{Asia Pacific Center for Theoretical Physics, Pohang, 37673, Republic of Korea}
	\affiliation{Department of Physics, Pohang University of Science and Technology, Pohang, 37673, Republic of Korea}
	
	\author{Youngjai~Park}                 
    \email{youngjai.park16@gmail.com}
	\affiliation{Center for Neuroscience Imaging Research, Institute for Basic Science, Suwon 16419, Republic of Korea}
	\affiliation{Sungkyunkwan University, Suwon 16419, Republic of Korea}

    \author{Jaesung~Choi}                \email{joseph9463@kias.re.kr}
	\affiliation{Center for Artificial Intelligence and Natural Sciences, Korea Institute for Advanced Study, Seoul 02455, Republic of Korea}
    
	\author{Younghwa~Cha}                       \email{youngcha1094@gmail.com}
	\affiliation{Center for Neuroscience Imaging Research, Institute for Basic Science, Suwon 16419, Republic of Korea}
	\affiliation{Sungkyunkwan University, Suwon 16419, Republic of Korea}
	\affiliation{Research Institute of Slowave Inc., Seoul 06160, Republic of Korea}
	\author{UnCheol Lee}
	\email{uclee@med.umich.edu}
	\affiliation{Department of  Anesthesiology, University of Michigan Medical School, Ann Arbor, Michigan, 48109, United States of America}

	\author{Joon-Young~Moon}                    \email{joon.young.moon@gmail.com (Corresponding author)}
	\affiliation{Center for Neuroscience Imaging Research, Institute for Basic Science, Suwon 16419, Republic of Korea}
	\affiliation{Sungkyunkwan University, Suwon 16419, Republic of Korea}
	
	\author{Jong-Min~Park}                      \email{jongmin.park@apctp.org (Corresponding author)}
	\affiliation{Asia Pacific Center for Theoretical Physics, Pohang, 37673, Republic of Korea}
	\affiliation{Department of Physics, Pohang University of Science and Technology, Pohang, 37673, Republic of Korea}

	\begin{abstract}
We analyze electroencephalography (EEG) signals using the ordinal pattern framework to investigate whether different human brain states can be distinguished based on the disorder of EEG dynamics.
Rather than analyzing raw EEG signals, we focus on the principal mode of EEG phase dynamics, reflecting anterior–posterior information flow, and quantify disorder using permutation entropy.
We apply this to two datasets: (i) EEG recordings from a general anesthesia protocol, and (ii) EEG recordings acquired in the resting state from healthy control subjects and individuals with inattentive-type attention deficit hyperactivity disorder (ADHD), including eyes-open and eyes-closed conditions.
We find that the permutation entropy distributions exhibit a clear dependence on brain state.
In particular, conscious, inattentive-type ADHD, and eyes-closed conditions show lower mean values and larger standard deviations of permutation entropy.
To evaluate the discriminative power of permutation entropy, we train classification models using permutation entropy as the input feature.
The results show that the distinction between conscious and unconscious states can be reliably captured in the general‑anesthesia dataset. In the resting‑state dataset, eyes‑open and eyes‑closed conditions are distinguishable, whereas classification between control and inattentive-type ADHD groups does not show clear separability.
This indicates that information not captured in ordinal patterns, such as the original time-series values, may play a more crucial role in detecting inattentive-type ADHD.
Our findings demonstrate that permutation entropy derived from EEG phase dynamics provides an effective indicator of brain states, particularly in relation to consciousness, while also highlighting its limitations for identifying individuals with inattentive-type ADHD.
	\end{abstract}

	\keywords{electroencephalography, brain states, ordinal patterns, permutation entropy, inattentive-type ADHD}

	\maketitle

	\section{Introduction}
Complex systems are composed of multiple interacting components whose mutual interactions give rise to emergent behaviors~\cite{mitchell2009complexity}.
The brain is a paradigmatic example of such a system, consisting of neural cells connected through a fractal network structure~\cite{smith2021neurons,reese2012analyzing}. These complex interactions generate rich emergent phenomena~\cite{gonzalez2007complex}, including perception, memory, and consciousness. The levels of these cognitive functions can vary with factors such as anesthesia~\cite{varley2020differential}, neurological disorders~\cite{yang2013mental}, development~\cite{frohlich2024sex}, and aging. A central challenge is to identify and characterize these functional states of the brain using experimentally accessible signals that reflect underlying neural activity. Understanding the relationship between brain states and the statistical characteristics of brain wave signals is therefore of significant importance.

Among the available modalities, electroencephalography (EEG) is widely used due to its non-invasive nature and high temporal resolution~\cite{gonzalez2019decreased,mateos2021using,schwarz2020analyzing}.
Despite its advantages, EEG signals are often contaminated by various artifacts, including subject motion, sweating, variations in skin conductance, and environmental factors such as humidity.
Together with technical noise arising from amplifiers and electrodes, these factors introduce significant noise into EEG datasets.

To mitigate the influence of such noise, ordinal-pattern-based analysis has been widely employed as a robust approach for time-series analysis. This framework analyzes a symbolic sequence of permutation orders, constructed not from the absolute values of a signal but from the relative ordering of neighboring data points~\cite{bandt2002permutation}. As long as the noise is not strong enough to alter the rank ordering of a few consecutive data points, the resulting permutation-order sequence remains unaffected, while still retaining essential information related to the underlying complexity of the data. In particular, the permutation entropy (PE), defined as the Shannon entropy of the ordinal pattern distribution, has been widely used as a quantitative measure of signal irregularity and complexity~\cite{smaal2021complexity}.

Previous research has shown that entropy-based measures of EEG signal complexity are powerful tools for investigating neurological and neuropsychiatric disorders~\cite{chu2017potential,jui2023application,hernandez2023brain,pallathadka2024investigating,al2022complexity,boaretto2023spatial,gu2024eeg, gancio2024permutation, catherine2022detection, niu2020comparing}. PE is shown to provide high accuracy in distinguishing attention deficit hyperactivity disorder (ADHD) subjects from control groups~\cite{catherine2022detection} and enhanced reliability in capturing topological information related to normal and disordered brain functioning~\cite{niu2020comparing}.
Along with the temporal PE, spatial PE, computed from the spatial arrangement of EEG electrodes, is found to differentiate eyes-open and eyes-closed resting states~\cite{boaretto2023spatial, gancio2024permutation}.

Despite these advantages, ordinal-pattern-based analysis faces intrinsic limitations when applied to high-dimensional time series. As data dimensionality increases, the number of possible ordinal patterns grows rapidly, making it increasingly difficult to define and interpret meaningful patterns. This issue is especially pronounced for EEG data, which consists of high-dimensional time series recorded simultaneously from many electrodes~\cite{deng2017multivariate,zeng2018characterizing,ma2021modified}. Consequently, determining how to appropriately construct ordinal patterns from multichannel EEG recordings becomes a nontrivial problem.
Therefore, identifying a representative low-dimensional time series that sufficiently captures the essential information embedded in high-dimensional EEG recordings would provide a more reliable and effective basis for applying ordinal-pattern-based analysis.

Recent studies have demonstrated that the phase dynamics of EEG signals provide critical insights into the directionality of information flow in the brain~\cite{Daffertshofer2018, alamia2019, zhang2018, alamia2023, mohan2024, Park2025.03.12.642768}.
By analyzing phase-lead and phase-lag relationships between EEG signals, we may infer the directionality of the information flow between different brain regions~\cite{stam2012,moon2015,alamia2023,mohan2024}.
At the system level, such phase-lead and phase-lag relationships are most meaningfully defined relative to the collective phase organization of the signals, rather than with respect to an arbitrary absolute reference. Research on phase dynamics~\cite{alamia2023,mohan2024,Park2025.03.12.642768} has identified two principal modes of information flow: top-down flow mode, where information propagates from higher-order cognitive regions to lower-order sensory areas, and bottom-up flow mode, where information moves in the reverse direction, from sensory regions to higher-order cognitive areas. In EEG and electrocorticography (ECoG) studies, top-down flow is typically characterized by anterior cortical signals phase-leading posterior cortical signals, whereas bottom-up flow is observed when posterior
signals phase-lead anterior signals~\cite{alamia2023,mohan2024,Park2025.03.12.642768}.
This structure allows high-dimensional multichannel EEG data to be effectively represented by the time series of a dominant principal mode.

In this paper, we apply an ordinal-pattern-based approach to analyze EEG signals. While previous studies have directly applied ordinal pattern analysis to EEG amplitude time series, we instead focus on the time series of the principal modes of the phase dynamics, which capture temporal fluctuations between top-down and bottom-up information flow modes. This approach is applied to two distinct datasets: (i) EEG recordings from subjects undergoing general anesthesia, in which seven distinct brain states are identified, and (ii) EEG recordings from both healthy control subjects and individuals diagnosed with the inattentive-type ADHD (hereafter abbreviated as inADHD).

We demonstrate that permutation entropy computed from the principal mode of EEG phase dynamics exhibits several advantageous properties compared to that obtained directly from raw EEG signals. In particular, we find that the phase-based permutation entropy converges with significantly shorter time-series lengths, allowing reliable estimation of ordinal pattern statistics from a smaller amount of data. Furthermore, it shows consistent behavior across different choices of the embedding dimension, indicating strong robustness.
This phase-dynamics representation reveals that the distribution of permutation entropy varies systematically with the level of consciousness, the presence of inADHD, and two resting-state conditions with eyes open and eyes closed. Specifically, higher levels of consciousness, individuals with inADHD, and eyes-closed states are characterized by lower mean values and higher standard deviations of permutation entropy, in contrast to their respective counterparts. This behavior remains robust across variations in both the time series segment length and the embedding dimension used to construct ordinal patterns.

In analogy, conventional raw EEG signals resemble individual waves on the ocean, high-dimensional, irregular, and locally fluctuating. In contrast, the principal mode of phase dynamics captures the tidal motion, representing a coherent, system-level pattern that governs the overall behavior of the ocean. Our results suggest that changes in brain state during anesthesia are better reflected in these global, low-dimensional dynamics than in local fluctuations of EEG. Thus, the distinction between conscious and unconscious states may be understood as a shift in the underlying global coordination of brain activity.
These findings demonstrate the potential of permutation entropy derived from EEG phase dynamics as a robust indicator of human brain states, such as anesthetic depth, which is a crucial factor in ensuring patient safety during clinical procedures~\cite{widman2000quantification}.

The paper is organized as follows. Section~\ref{sec:three} describes the datasets, the principal mode of EEG phase dynamics, and the permutation entropy. Section~\ref{sec:four} presents the analysis and discussion. Finally, Section~\ref{sec:five} provides concluding remarks and discusses the implications of our findings.

	\begin{figure*}
        \includegraphics[scale=0.52]
        {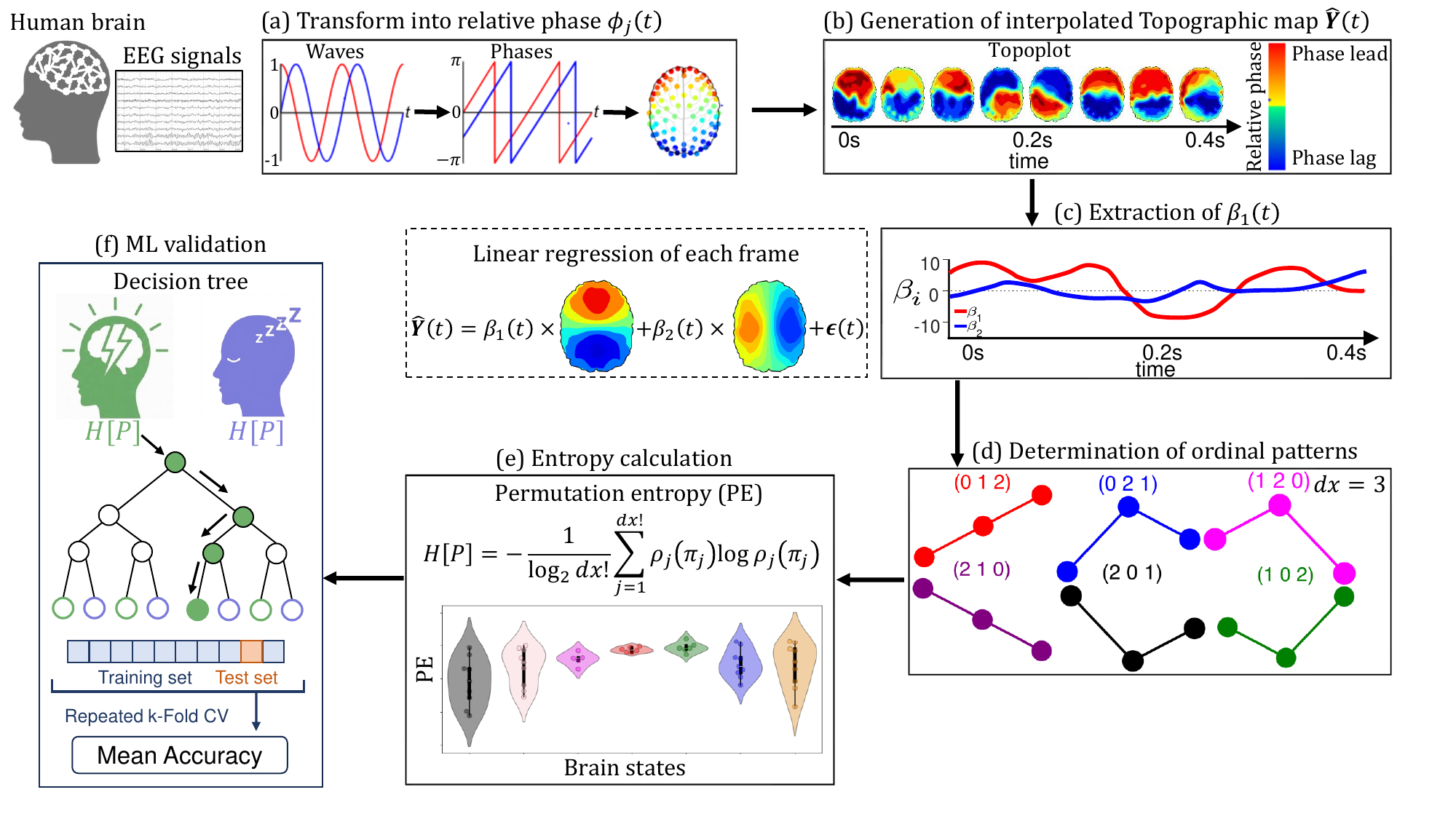}
		\caption{\textbf{Workflow diagram}: (a) transform EEG signals into phase dynamics; (b) generate a topographic map via spatial interpolation; (c) extraction of the time series $\beta_1(t)$ via linear regression; (d) construction of ordinal patterns from $\beta_1(t)$ for a given embedding dimension $d x$; (e) computation of the normalized permutation entropy $H[P]$ from the ordinal pattern distribution $\{\rho(\pi_j)\}$; and (f) validation of distinguishing power using decision-tree-based classifiers.
        }
		\label{fig:workflow_pipeline}
	\end{figure*}
    
    \section{Methodology}
	\label{sec:three}

In this study, we analyze two independent EEG datasets to investigate the degree of disorder in the directional information flow of human brain dynamics across distinct brain states:

(i) \textbf{Dataset~\hyperref[sec:two_a]{I}}: It consists of EEG recordings obtained from nine healthy adult volunteers (aged 20-40 years) undergoing a controlled general anesthesia protocol.
These participants exhibit up to seven distinct brain states: eyes-closed (EC), propofol injection period (P), loss of consciousness (LOC), burst period (B), suppression period (S), post-burst/suppression anesthesia (PBS), and recovery of consciousness (ROC)~\cite{Park2025.03.12.642768}. EEG recordings were obtained using a 128-channel system. We note that not all participants exhibit all seven brain states mentioned above. In particular, the number of participants displaying each state is nine for EC, P, PBS, and ROC states, six for B and S states, and five for the LOC state. Please refer to Appendix~\ref{sec:two_a} for more details on the general anesthesia dataset.

(ii) \textbf{Dataset~\hyperref[sec:two_b]{II}}: It consists of resting-state EEG data from 106 participants: 40 with inADHD and 66 healthy controls (aged $\geq$11 years). EEG data were recorded under alternating eyes-open (EO) and eyes-closed (EC) conditions, using a 128-channel setup.
Accordingly, we have four distinct states: Control\_EO, Control\_EC, inADHD\_EO, and inADHD\_EC, corresponding to eyes-open and eyes-closed conditions for the healthy Control and inADHD groups, respectively.
Please see Appendix~\ref{sec:two_b} for more details on the inADHD dataset. 

First, we perform the following data preprocessing steps: (i) notch filtering at 59-61 Hz to remove power line noise; (ii) bandpass filtering (0.5-100 Hz) to exclude slow drifts and high-frequency artifacts; (iii) bad channel rejection based on voltage outliers (0.0001-100~$\mu$V); and (iv) alpha-band (8–12 Hz) filtering to extract the alpha-band signal. Finally, relative phase signals were computed and whole-brain topographic maps were constructed by averaging the signals within 100 ms time windows, followed by regression analysis to improve computational efficiency and robustness, as described in detail below. Further details on the data preprocessing are provided in Appendix~\ref{sec:edap} and reference~\cite{Park2025.03.12.642768}.
    
    We then proceed to analyze the EEG signals recorded from the $N = 128$ channels through a series of steps as illustrated in Fig.~\ref{fig:workflow_pipeline}. We represent the raw data as a series of 128-dimensional vectors, $\boldsymbol{y}(t) = (y_1(t), y_2(t), \dots, y_{128}(t))$, where  $y_j(t)$ is the signal recorded from channel $j$ at time $t$.
    By using the Hilbert transform, we decompose the data into amplitude and phase, where the phase is denoted by $\boldsymbol{\theta} (t)$.
    The amplitude represents the magnitude of the signal, while the phase represents its relative temporal position.
    
    To characterize phase dynamics at the system level, we describe instantaneous phases relative to the mean phase across EEG channels, following the approach used in~\cite{Park2025.03.12.642768}, which provides a common system-level reference. This procedure yields a time-resolved representation of relative phase patterns that summarizes how phases are organized across channels. From this representation, we extract the principal mode of the phase dynamics, whose time series is used as the input for subsequent ordinal pattern and permutation entropy analyses.
    
    More precisely, the relative phase $\boldsymbol{\phi}(t)$ is defined as:
\begin{equation}
    e^{i \phi_j(t)} = e^{i (\theta_j(t) - \Omega (t))},
\end{equation}
with respect to the global mean phase $\Omega(t)$ computed from
\begin{equation}
    R e^{i \Omega (t)} = \frac{1}{N} \sum_{j=1}^N e^{i \theta_j (t)},
\end{equation}
with the real-valued global amplitude $R$. 
Then, we perform dimensionality reduction to focus on macroscopic brain patterns. 

    For each 100 ms window, we first compute relative phase signals from 128 EEG channels and spatially interpolate them to obtain a whole-brain relative phase topographic map, as presented in Fig.~\ref{fig:workflow_pipeline}(b).
    Following Ref.~\cite{Park2025.03.12.642768}, we use two dominant regressors $\boldsymbol{X}_1$ and $\boldsymbol{X}_2$ derived from $K$-means centroids, each representing anterior-posterior directionality (anterior part either phase-leading or lagging the posterior part), and left-right hemisphere directionality (left hemisphere phase-leading or lagging the right hemisphere)  (see also Appendix~\ref{sec:two_d} for details).
    We note that although we have employed $K$-means clustering to extract these dominant modes, the resulting modes are almost identical to the principal modes obtained via principal component analysis (PCA).
    
    We then perform multiple linear regression of the vectorized topographic map $\hat{\boldsymbol{Y}}(t)$ onto $\boldsymbol{X}_1$ and $\boldsymbol{X}_2$ by minimizing the residual term $\boldsymbol{\epsilon}(t)$.
    To capture the temporal evolution of the data along these principal modes, we perform multiple linear regression of the topographic map $\hat{\boldsymbol{Y}}(t)$ at time $t$ as follows:
    \begin{equation}
        \hat{\boldsymbol{Y}}(t) = \beta_1 (t) \boldsymbol{X}_1 + \beta_2 (t) \boldsymbol{X}_2 + \boldsymbol{\epsilon}(t), ~\label{eq:sb}
    \end{equation}
    where $\beta_1(t)$ and $\beta_2(t)$ are the time-resolved regression coefficients of the two fixed centroid-based regressors, and $\boldsymbol{\epsilon}(t)$ represents the residual.
    
    Accordingly, $\beta_1(t)$ and $\beta_2(t)$ quantify the time-varying contributions of the anterior-posterior and left-right phase patterns, respectively. For example, $\beta_1(t)$ becomes positive when anterior regions are phase-leading relative to posterior regions, and negative when the direction is reversed. Please see Appendix~\ref{app:mb} and Ref.~\cite{Park2025.03.12.642768} for details on $\beta_1(t)$ and $\beta_2(t)$. Previous research has demonstrated that the degree of anterior-to-posterior directionality correlates with brain state dynamics~\cite{lee2013disruption,moon2017structure}.
    Further details on the procedure for obtaining $\beta_1$ are provided in the tutorial code included in the Supplementary Material.

    To analyze the macroscopic states of human brain activity, we quantify the degree of disorder in the EEG phase dynamics. A standard measure of disorder is Shannon entropy~\cite{shannon1948mathematical}. However, estimating Shannon entropy from a continuous-valued time series $x(t)$ is challenging, as it requires accurate estimation of the probability distribution $\mathcal{P}(x)$. To circumvent this difficulty, we adopt a symbolic approach based on the ordinal patterns of the time series.
    
   The symbolic sequence is obtained by mapping consecutive data points in the time series into a corresponding ordinal pattern. The number of data points used in this mapping is referred to as the permutation order or embedding dimension, denoted by $dx$. For a given $dx$, the total number of possible ordinal patterns is given by $dx!$.
    As a simple example, we consider the case $dx=2$. We first compare the order of two consecutive data points $x (t_i)$ and $x (t_{i+1})$ for $i=1, 2, \dots, M-1$ , where $M$ denotes the data length. Then, we assign each of the $2!$ possible ordinal patterns as $\pi_1=(0,1)$ for $x (t_i) < x (t_{i+1})$ or
$\pi_2=(1,0)$ for $x (t_{i+1}) < x (t_i)$.
Applying this process over all  $i$, we obtain a symbolic sequence of ordinal patterns and the probability $\rho(\pi_j)$ of pattern $\pi_j$ for $j=1$ and $2$. The degree of disorder in such an ordinal pattern sequence is quantified by the \textit{permutation entropy}~\cite{bandt2002permutation,pessa2021ordpy}, which is defined as
\begin{equation}
    S[P] = -\sum_{j=1}^{dx!} \rho(\pi_j) \log_2 \rho(\pi_j),
\end{equation}
with $P=\{\rho(\pi_j)\}$.
For consistent comparison across different values of $dx$, we use the normalized permutation entropy, defined as
\begin{equation}
    H[P] = \frac{S[P]}{\log_2 dx!}, \label{eqhp}
\end{equation}
which is bounded between 0 and 1 regardless of the choice of $dx$. The detailed explanation of the steps used in computing ordinal patterns and permutation entropy is given in Appendix~\ref{sec:pee}.

    \begin{figure}
\includegraphics[width=0.45\textwidth]
{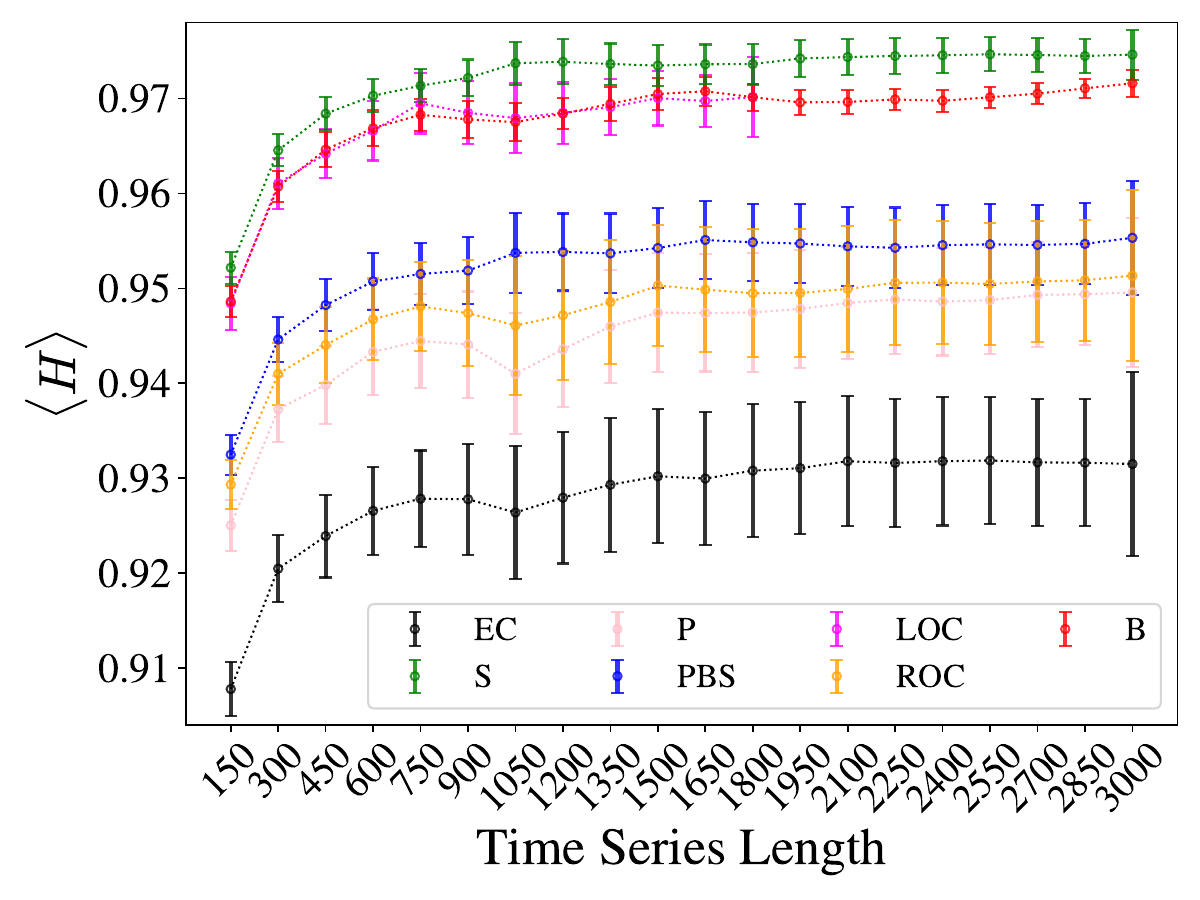}
\vskip -0.1in
\caption{\textbf{Time-series length dependence of mean $\beta_1$-PE $\langle H\rangle$ across different brain states in the general anesthesia dataset~\hyperref[sec:two_a]{I}:} Symbols represent the mean $\beta_1$-PE $\langle H \rangle$ computed using $d x = 4$ over time segments for the following brain states: eyes closed (EC; black), propofol injection period (P; pink), loss of consciousness (LOC; magenta), burst period (B; red), suppression period (S; green), post-burst/suppression anesthesia (PBS; blue), and recovery of consciousness (ROC; orange).}
		\label{fig:beta1_pe_time_length}
	\end{figure}

    \begin{figure}
\includegraphics[width=0.45\textwidth]{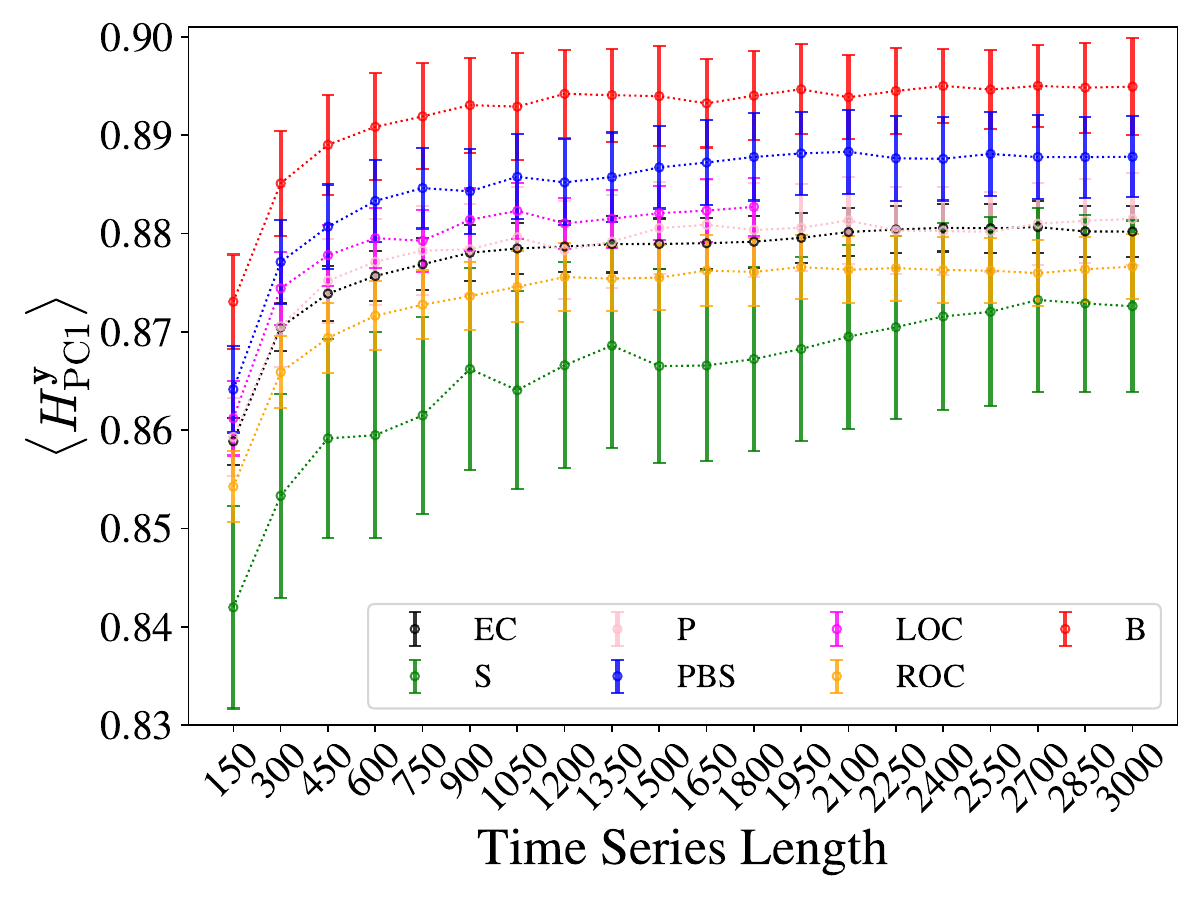}
\vskip -0.1in
\caption{
\textbf{Time-series length dependence of mean permutation entropy computed from raw EEG data in the general anesthesia dataset~\hyperref[sec:two_a]{I}:} Symbols represent the mean permutation entropy $\langle H^{\boldsymbol{y}}_{\mathrm{PC}1} \rangle$ computed from the first principal component in the raw EEG data using $d x = 4$. Brain states are represented by the same colors as in Fig.~\ref{fig:beta1_pe_time_length}.
}
	\label{fig:raw_pe_time_length}
	\end{figure}

	\section{Results}
	\label{sec:four}
    It is expected that the degree of disorder in human brain activity is significantly influenced by the level of consciousness or the presence of inADHD.
    To reveal characteristic features due to these influences, we compute the normalized permutation entropy $H$ of the anterior-posterior directionality pattern $\beta_1(t)$, referred to as $\beta_1$-PE,
    for each brain state in the two datasets~\hyperref[sec:two_a]{I} and ~\hyperref[sec:two_b]{II}.

    \subsection{General anesthesia}
\begin{figure*}
\includegraphics[width=\textwidth]
{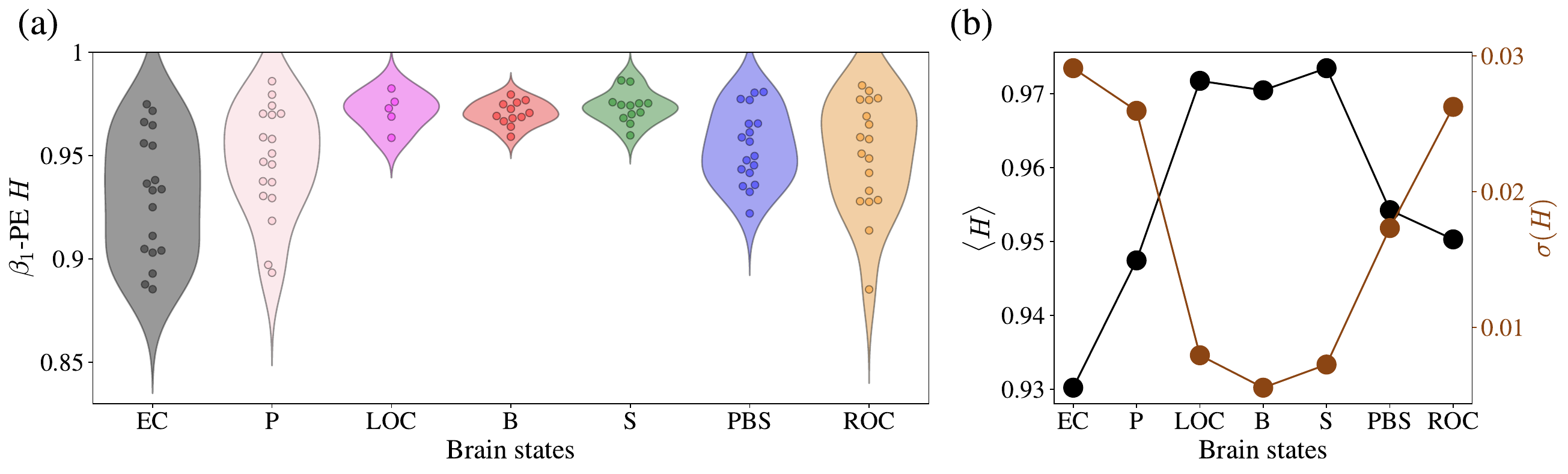} 
\vskip -0.1in
\caption{\textbf{Statistical properties of $\beta_1$-PE ($H$) obtained from the general anesthesia dataset~\hyperref[sec:two_a]{I}:}
(a) Violin plots and symbols represent the distribution of $\beta_1$-PE for time segments of length 1500 with $d x = 4$ across subjects. Brain states are represented by the same colors as in Fig.~\ref{fig:beta1_pe_time_length}.
(b) Black and brown symbols indicate the mean and standard deviation of $\beta_1$-PE computed over the time segments, respectively.}
		\label{fig:beta1_pe_distribution}
	\end{figure*}

\begin{figure}[b]
\includegraphics[width=0.45\textwidth]{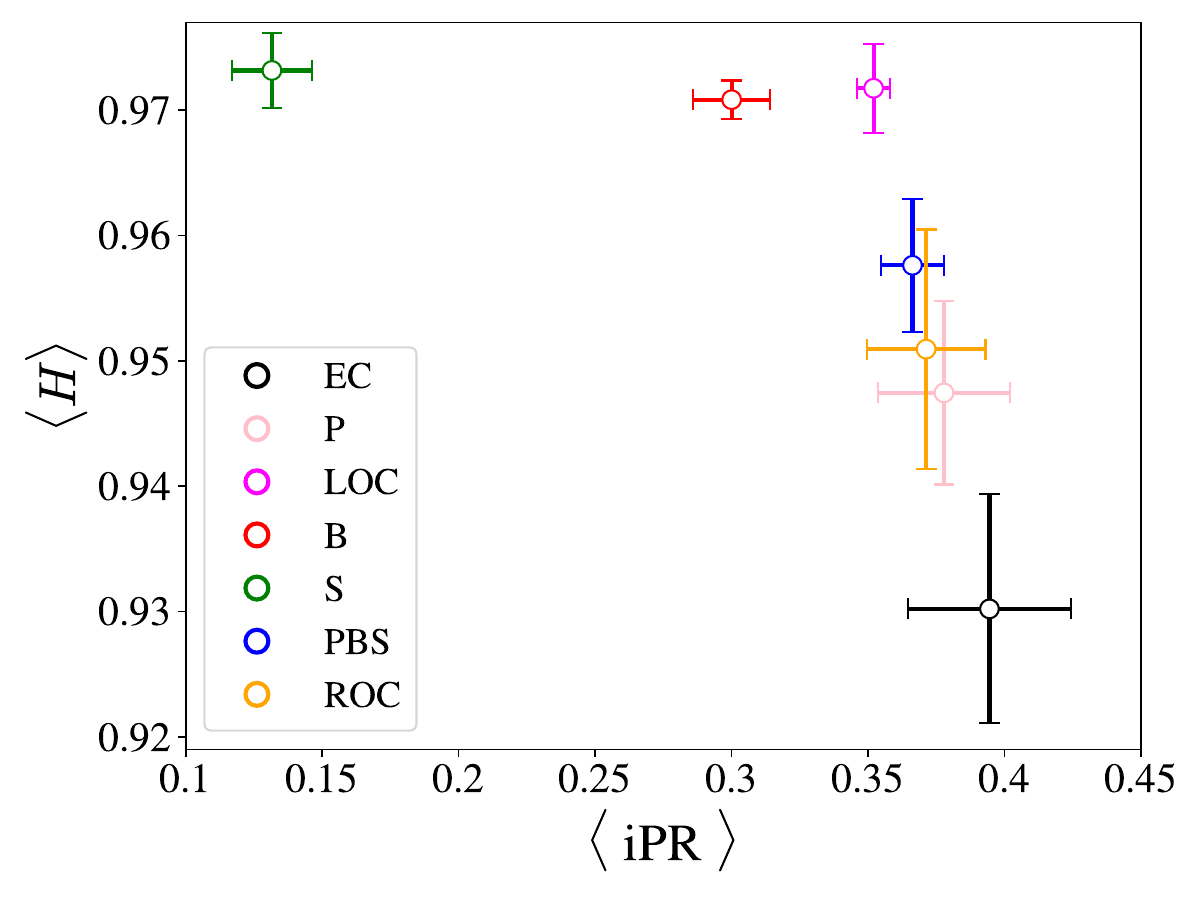}
\vskip -0.1in
\caption{\textbf{Correlation between $\beta_1$-PE and the inverse participation ratio (iPR):} Both $\beta_1$-PE and the iPR are computed for the same time segments as those used in Fig.~\ref{fig:beta1_pe_distribution}. Error bars represent the standard errors. Brain states are represented by the same colors as in Fig.~\ref{fig:beta1_pe_time_length}. 
        }
		\label{fig:beta1_pe_ipr_correlation}
	\end{figure}

We first examine how the $\beta_1$-PE ($H$) varies with the length of the $\beta_1(t)$ time series across different embedding dimensions for each brain state across all subjects in the general anesthesia dataset~\hyperref[sec:two_a]{I}. We divide the $\beta_1 (t)$ time series into several segments of a given length for which $H$ is computed.
To minimize information loss, when the segment length is longer than half of the total time-series length, we use two overlapping segments: one starting from the initial point and the other ending at the final point of the time series.
The mean value $\langle H\rangle$ computed with $d x = 4$
as a function of segment length
for different brain states in dataset~\hyperref[sec:two_a]{I}
is presented in Fig.~\ref{fig:beta1_pe_time_length}.
For all seven brain states, $\langle H\rangle$ saturates to a constant value once the time-series length exceeds a certain threshold.
In particular, for $dx=4$, this convergence occurs at a time series length of $1500$, which we use in the following analyses.

We also examine the dependence of the results on the embedding dimension $dx$, as shown in Fig.~\ref{fig:dx_variation_time_length}. 
The results show that the overall behavior remains consistent across different values of $dx$ with differences only in error bar magnitude and saturation points.
This indicates that $\beta_1$-PE is robust with respect to the choice of embedding dimension $dx$. Additional analysis of the robustness of $\beta_1$-PE with respect to $d x$ is provided in Appendix~\ref{sec:extra}.

To clarify the advantage of employing phase dynamics, we compare these results with those obtained by applying the same analysis directly to the original EEG time series.
Figure~\ref{fig:raw_pe_time_length} shows the permutation entropy $H^{\boldsymbol{y}}_{\mathrm{PC}1}$ computed with $dx=4$ from the first principal component obtained via PCA of the raw data $\boldsymbol{y}(t)$.
Compared to $\beta_1$-PE, the permutation entropy from the raw data requires a much longer time-series length to reach saturation, indicating that a larger number of data points is required to avoid finite-sample bias in the estimation of ordinal pattern statistics.
More importantly, the raw-data-based permutation entropy exhibits a strong dependence on the embedding dimension $d x$, as shown in Fig.~\ref{fig:raw_dx_variation_time_length}.
In particular, even the relative ordering of permutation entropy across brain states varies with the choice of $d x$.
These results indicate that the permutation entropy obtained from the principal component of the raw EEG signal is strongly influenced by the choice of embedding parameter rather than reflecting intrinsic properties of the brain state. 
In contrast, $\beta_1$-PE based on phase dynamics preserves state-dependent information and exhibits consistent behavior across brain states.
Further analysis of the limitations of the raw-data-based permutation entropy is provided in Appendix~\ref{sec:review}.

Figure~\ref{fig:beta1_pe_distribution}(a) presents violin plots of the $\beta_1$-PE ($H$) for each brain state in dataset~\hyperref[sec:two_a]{I}. The distributions indicate that the $\beta_1$-PE distributions for the conscious states, EC and ROC, are more broadly distributed and have lower mean values than those for the unconscious states, LOC, B, and S.
The $\beta_1$-PE distributions of the boundary states, P and PBS, exhibit characteristics intermediate between these two regimes.

To further quantify this observation, we compute the mean $\langle H\rangle$ and standard deviation $\sigma(H)$ of $\beta_1$-PEs
for each brain state, as shown in Fig.~\ref{fig:beta1_pe_distribution}(b). The results confirm that anesthetized states (LOC, B, and S) exhibit higher mean entropy and smaller standard deviation,
indicating increased disorder and reduced inter-subject variability.

We also perform the same analysis for the second principal mode $\beta_2(t)$ and find qualitatively similar behaviors, including comparable convergence with increasing time-series length, robustness with respect to the embedding dimension $dx$, a consistent correlation with the level of consciousness, and an anti-correlation between the mean and standard deviation. 
However, since $\beta_1(t)$ and $\beta_2(t)$ are strongly correlated, incorporating $\beta_2(t)$ does not provide additional independent information. Therefore, focusing on $\beta_1(t)$ alone is sufficient to characterize the brain states. Detailed results for $\beta_2(t)$ are provided in Appendix~\ref{sec:beta2}.

We now relate the mean $\beta_1$-PE $\langle H\rangle$ to a quantitative measure of the level of consciousness. For this purpose, we use the inverse participation ratio (iPR) index defined as~\cite{eriksen2003modularity,beugeling2015global}
    \begin{equation}
		\textrm{iPR}= \sum_i \left( \frac{\lambda_i}{\sum_j \lambda_j} \right)^2,\label{eq:ipr}
	\end{equation}
where $\lambda_i$ denotes the eigenvalue of the EEG channel-channel covariance matrix, obtained via singular value decomposition. The participation ratio quantifies the heterogeneity of a given distribution by measuring the proportion of principal components that contribute significantly to the overall variance.
Thus, the higher the value of iPR, the lower the depth of anesthesia~\cite{Park2025.03.12.642768}.

We observe a negative correlation between mean $\langle H \rangle$ and $\langle \textrm{iPR} \rangle$ index, as shown in Fig.~\ref{fig:beta1_pe_ipr_correlation}. This negative correlation
indicates a strong relation between $\beta_1$-PE and the level of consciousness.
These findings suggest that $\beta_1$-PE can serve as a key feature for distinguishing between conscious and unconscious states.

\begin{figure}
\includegraphics[width=0.45\textwidth]
{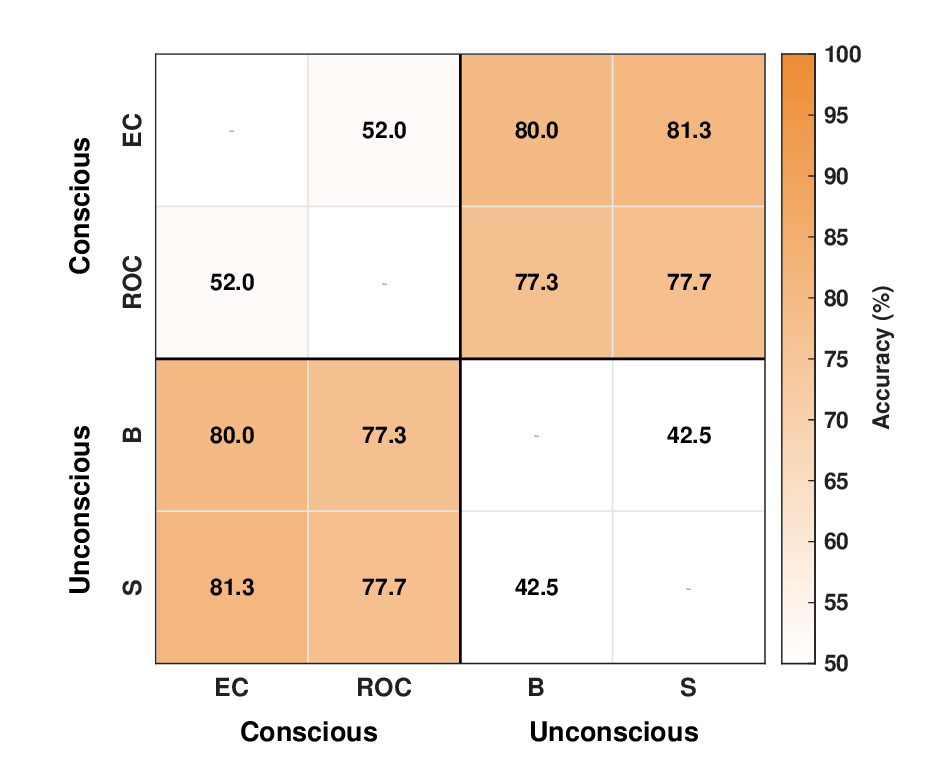}
\vskip -0.1in
\caption{\textbf{Classification accuracy of a decision tree for distinguishing conscious and unconscious states in the general anesthesia dataset~\hyperref[sec:two_a]{I}:} The heatmap indicates the classification accuracy (\%) of a classifier trained using $\beta_1$-PE as a single feature. The evaluation is performed for two conscious states, eyes closed (EC) and recovery of consciousness (ROC), and two unconscious states, burst (B) and suppression (S).}
\label{fig:classification_anesthesia}
	\end{figure}

To assess the performance of $\beta_1$-PE in a discrimination task, we evaluate the classification accuracy obtained when $\beta_1$-PE is used as the input feature for a classifier.
A similar approach was adopted in a previous study~\cite{gancio2024permutation} to distinguish eyes-open and eyes-closed states, where a random forest classifier was employed.
In this study, we employ a decision tree model, which can be regarded as the basic unit of a random forest, to mitigate the risk of overfitting arising from limited data and features.
The model is validated through a 5-fold cross-validation (CV) repeated 10 times.
The resulting pairwise binary classification accuracies for each pair of brain states are presented in Fig.~\ref{fig:classification_anesthesia}.
Here, we exclude the boundary states (P and PBS), whose ground-truth classification as conscious or unconscious is ambiguous. Additionally, we exclude the LOC state to avoid sample-size bias given its small sample size.
These results demonstrate that brain states with different levels of consciousness are readily distinguishable, indicating that $\beta_1$-PE is an effective feature for distinguishing conscious from unconscious states.

	\subsection{Healthy controls and Inattentive-type ADHD across eyes-open and eyes-closed resting states}
	\label{sec:A2}
    \begin{figure}
\includegraphics[width=0.45\textwidth]
{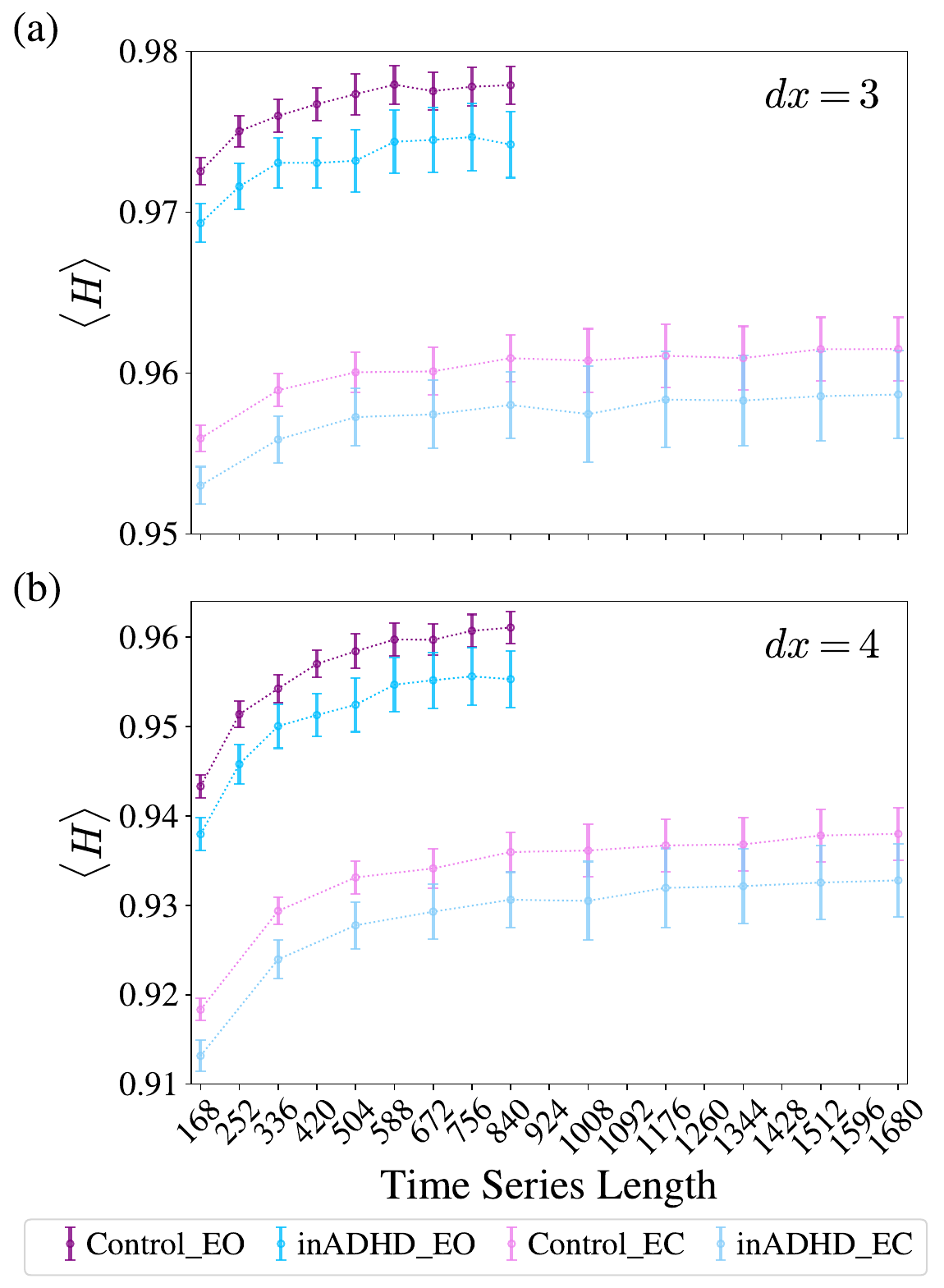}\\
\vskip -0.1in
\caption{\textbf{Time-series length dependence of $\beta_1$-PE across different brain states in the inADHD dataset~\hyperref[sec:two_b]{II}:} Symbols represent the mean
$\beta_1$-PE $\langle H \rangle$ computed over time segments for eyes-open and eyes-closed conditions for healthy controls (Control\_EO/Control\_EC; purple/light purple) and inattentive-type ADHD subjects (inADHD\_EO/inADHD\_EC; cyan/light cyan) using (a) $d x = 3$, and (b) $d x = 4$.
}
		\label{fig:beta1_pe_time_length_adhd}
	\end{figure}

	\begin{figure*}
\includegraphics[width=1\textwidth]
{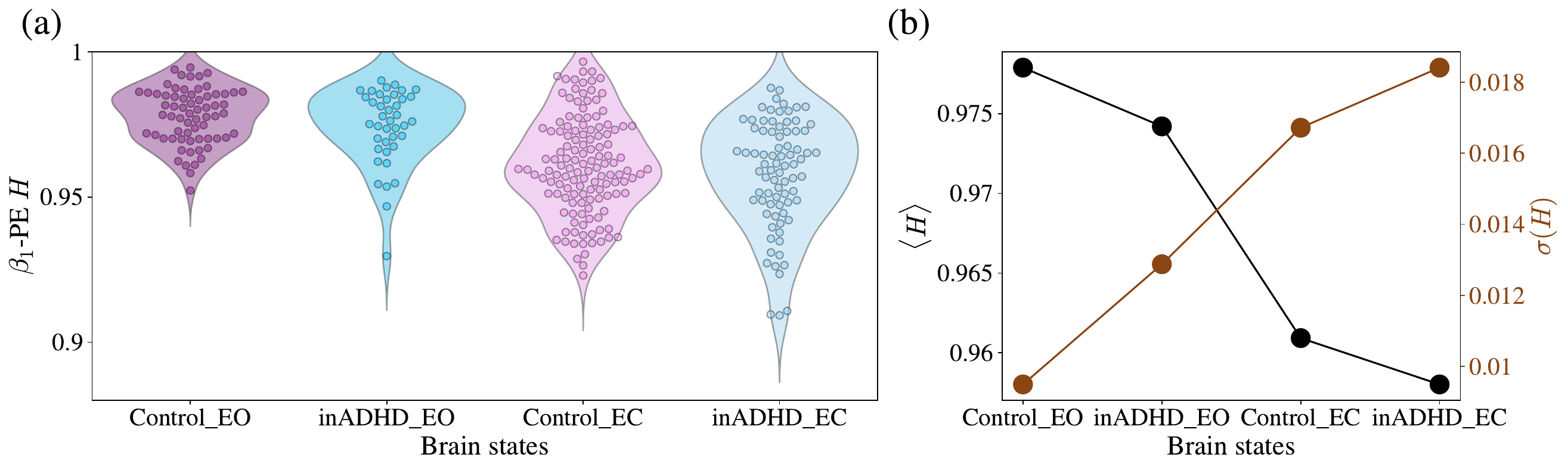}
\caption{\textbf{Statistical properties of $\beta_1$-PE obtained from the inADHD dataset~\hyperref[sec:two_b]{II}:}
(a) Violin plots and symbols represent the distribution of $\beta_1$-PE for time segments with length $840$ with $d x = 3$. The same colors represent the brain states as in Fig.~\ref{fig:beta1_pe_time_length_adhd}.
(b) Black and brown symbols indicate the mean and standard deviation of $\beta_1$-PE computed over time segments, respectively.}
		\label{fig:beta1_pe_distribution_adhd}
	\end{figure*}

    \begin{figure}
\includegraphics[width=0.5\textwidth]
{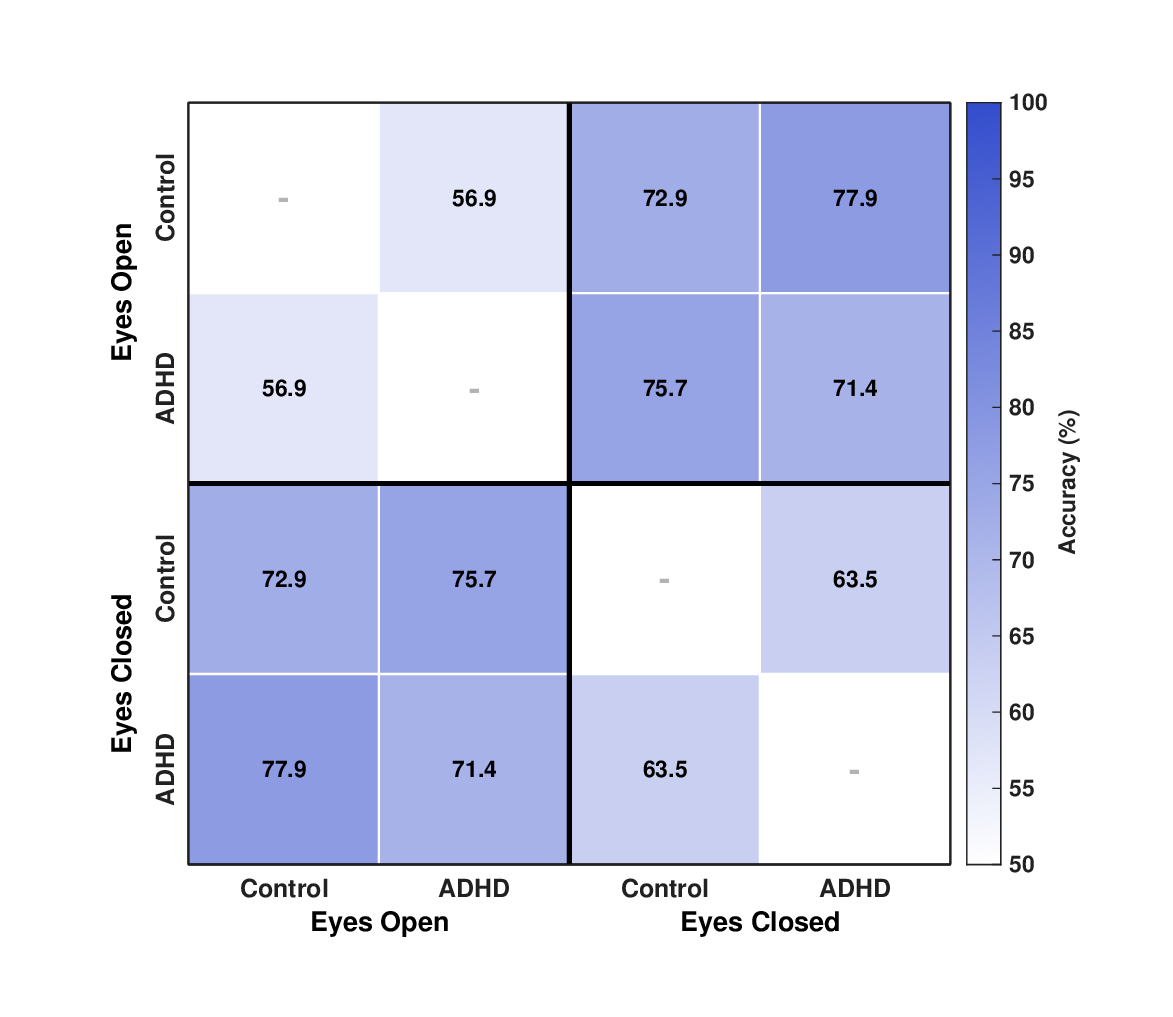}
\vskip -0.1in
\caption{\textbf{Classification accuracy of a random forest for distinguishing resting states and inADHD condition in the inADHD dataset~\hyperref[sec:two_b]{II}:} The heatmap indicates the classification accuracy (\%) obtained using a classifier trained with $\beta_1$-PE as a single feature. The evaluation is performed for eyes-open and eyes-closed conditions in healthy controls (Control) and inattentive-type ADHD subjects (inADHD).}
\label{fig:classification_adhd}
	\end{figure}

We now apply the same approach to dataset~\hyperref[sec:two_b]{II} to investigate differences associated with the presence of the inADHD condition as well as across eyes-open and eyes-closed resting states. As in dataset \hyperref[sec:two_a]{I}, we examine how the $\beta_1$-PE varies with the time-series length. As shown in Fig.~\ref{fig:beta1_pe_time_length_adhd}, saturation of $\langle H \rangle$ occurs at approximately $840$ points for $dx = 3$, whereas no apparent saturation is observed for $dx = 4$. Accordingly, we adopt $dx=3$ and a time-series length of $840$ for the subsequent analyses.

Figure~\ref{fig:beta1_pe_distribution_adhd}(a) displays the violin plots of $\beta_1$-PE for healthy control and inADHD groups across EO and EC states. We observe that the shape of the distribution differs significantly across brain states.
The standard deviation $\sigma(H)$ of the $\beta_1$-PE is consistently higher in inADHD subjects than in the control group in both EO and EC conditions. In addition, the mean $\beta_1$-PE $\langle H\rangle$ shows lower values for inADHD individuals compared to healthy controls in both conditions.
The elevated standard deviation may reflect the characteristic attentional instability of individuals with inADHD, manifesting as greater fluctuations in anterior-posterior information processing capacity.
Moreover, both groups exhibit increased $\sigma(H)$ and decreased $\langle H \rangle$ during EC states compared to EO states.

The finding that mean $\beta_1$-PE is lower in the case of inADHD patients than in healthy control subjects is consistent with those of several previous studies on different types of ADHD.
Previous research on different types of entropy measures in ADHD subjects indicates reduced multiscale entropy in ADHD patients than healthy controls~\cite{guan2023complexity}, lower sample entropy for adult patients with ADHD in functional magnetic resonance imaging (fMRI) signals in resting-state~\cite{sokunbi2013resting}, decreased fuzzy entropy for ADHD patients in the magnetoencephalographic (MEG) activity signals~\cite{monge2015meg}, reduced approximate entropy in resting-state EEG of ADHD children~\cite{chen2019eeg}, reduced approximate entropy in ADHD adolescent boys during a continuous performance test (CPT)~\cite{sohn2010linear}, diminished EEG complexity in alpha frequency bands in ADHD during multi-source interference tasks~\cite{chenxi2016complexity}, reduced approximate, sample, and Shannon entropy in EEG signals of ADHD adults during 3-minute eyes-open and eyes-closed conditions~\cite{kaur2019}, and decreased permutation entropy in combined-type ADHD~\cite{ccetin2022case}.

Since dataset \hyperref[sec:two_b]{II} provides a sufficient number of permutation entropy values, we adopt the random forest classifier to assess classification performance metrics.
The accuracy of the random forest classifier across all pairwise combinations of the brain states in dataset \hyperref[sec:two_b]{II}, classified by both the resting states (eyes-open and eyes-closed) and the presence of inADHD, is shown as a heatmap in Fig.~\ref{fig:classification_adhd}. The results indicate that the brain states are clearly distinguished by eyes-open and eyes-closed resting states, whereas the presence of inADHD is not reliably detected by the classifier. This implies that the $\beta_1$-PE distributions of healthy controls and individuals with inADHD largely overlap, even though their means and standard deviations appear different. Taken together, these observations suggest that the differences in the statistics may be influenced by a few extreme values.
This further suggests that the information discarded during ordinal-pattern transformation may be crucial for detecting inADHD.
For example, our previous study~\cite{Park2025.03.12.642768} showed that the presence of inADHD can be distinguished by the duration or frequency with which $\beta_1(t)$ takes positive or negative values.
Thus, the inADHD condition is more readily identified using the value of $\beta_1$ itself, rather than through its relative temporal changes.

To examine the effect of the presence of inADHD on classification performance, we apply the classifier separately to datasets containing only the control group
or only the inADHD group.
Table~\ref{tab:classification_performance} shows that the classification metrics do not exhibit a substantial dependence on the presence of inADHD, except for the specificity, which shows a noticeable reduction.
The reduced specificity may be related to the smaller sample size.

\begin{table*}[ht]
    \centering
    \caption{Classification performance of the random forest model for eyes-open (EO) vs. eyes-closed (EC) across three data scenarios: All (Control+inADHD), Control Only, and inADHD Only. Here, EC is treated as the positive class when computing precision, recall, and F1 score (specificity corresponds to the true-negative rate for EO). Values represent mean $\pm$ standard deviation derived from repeated 10-Fold Cross-Validation.}
\label{tab:classification_performance}
    \begin{tabular}{lccccc}
        \hline\hline
        \textbf{Scenario} & \textbf{Accuracy} & \textbf{F1 Score} & \textbf{Precision} & \textbf{Recall} & \textbf{Specificity} \\
         & (\%) & (\%) & (\%) & (\%) & (\%) \\
        \hline
        All (Control+inADHD) & $71 \pm 1$  & $79 \pm 1$  & $78 \pm 1$ & $81 \pm 2$  & $53 \pm 3$ \\
        Control Only       & $73 \pm 1$  & $80 \pm 1$  & $80 \pm 1$ & $81 \pm 1$ & $58 \pm 3$ \\
        inADHD Only          & $71 \pm 3$ & $80 \pm 3$ & $76 \pm 1$ & $87 \pm 3$ & $41 \pm 4$ \\
        \hline
    \end{tabular}
\end{table*}

	\section{Conclusions}
	\label{sec:five}

Our findings demonstrate that permutation entropy applied to EEG phase dynamics, referred to as $\beta_1$-PE, provides a more efficient, robust, and discriminative characterization of human brain states compared to permutation entropy computed directly from raw EEG signals.
By focusing on a dominant anterior–posterior component of the EEG phase dynamics, the proposed approach captures system-level organization of phase relationships across channels rather than local signal magnitude, leading to a more physically meaningful representation of brain-state-dependent dynamics.

In particular, $\beta_1$-PE enables reliable estimation of ordinal pattern statistics from shorter time-series data, exhibits consistent behavior across different choices of the embedding dimension, and shows a clear systematic relationship with brain states.
Unconscious states, eyes-open conditions, and healthy controls are characterized by higher mean $\beta_1$-PE values, whereas their respective counterparts exhibit relatively lower mean values.
Across datasets, we consistently find that the standard deviation of $\beta_1$-PE distributions is inversely related to their mean values, indicating a nontrivial dependency of $\beta_1$-PE distributions on brain states.

To further quantitatively evaluate how effectively $\beta_1$-PE can distinguish between different brain states, we assess standard classification metrics using decision-tree-based classifiers. The results demonstrate that $\beta_1$-PE is a reliable feature for distinguishing consciousness levels and resting states (eyes-open and eyes-closed) with consistently high accuracy.
Specifically, the classification accuracy for distinguishing levels of consciousness or resting states is approximately 70–80\%, which is comparable to previous results obtained from classifications using a single feature~\cite{7848476,gancio2024permutation}. Importantly, $\beta_1$-PE is derived from a one-dimensional time series extracted in a purely data-driven manner without relying on intuitive assumptions or averaging across electrodes. These results indicate that comparable discrimination performance is achieved using a purely data-driven quantity derived from EEG phase dynamics.

In contrast, the presence of inADHD cannot be reliably detected by classifiers based on $\beta_1$-PE, despite apparent differences in the mean and standard deviation of $\beta_1$-PE distributions.
Notably, when considering only the mean values of $\beta_1$-PE computed for the inADHD and healthy control groups, the two groups appear clearly separated, with error bars showing little overlap. However, the classification results reveal that $\beta_1$-PE does not serve as an effective feature for discriminating between these two groups. This apparent discrepancy suggests that the $\beta_1$-PE distributions deviate significantly from Gaussianity and that the observed differences in the mean values are predominantly driven by a small number of extreme events rather than by systematic shifts of the central part of the distribution. This observation highlights that conclusions drawn solely from comparisons of mean values and associated error bars may be misleading, underscoring the importance of distribution-level and classification-based analyses when assessing the discriminative power of candidate features.
Therefore, complementary measures that explicitly capture such extreme events or distributional asymmetries may be necessary for reliably detecting the presence of inADHD or other neurodevelopmental conditions. This examination will be left for future work, as it requires larger and higher-quality datasets.

From a practical perspective, the development of robust indicators to assess the depth of anesthesia is of considerable importance~\cite{widman2000quantification,Jordan2008Anesthesiology}.
We demonstrate that the mean $\beta_1$-PE shows a clear correlation with iPR, an established indicator of consciousness, suggesting the potential of $\beta_1$-PE to serve as an effective feature for assessing consciousness levels.
Application of the present framework to larger and more diverse datasets may facilitate the development of robust, quantitative indicators of brain states in clinical settings.

As a direction for future work, it would be valuable to investigate the role of linear and nonlinear temporal correlations in the observed brain-state-dependent behavior, for example by combining ordinal pattern analysis with surrogate time series~\cite{zunino2017detecting,zunino2024revisiting}. This may help to elucidate the mechanisms underlying the differences captured by the proposed measure.

	\section*{Authors' contributions}
	{\noindent}
	Athokpam Langlen Chanu: Investigation, Methodology, Software, Validation, Visualization, Writing – original draft, Writing – review and editing. Youngjai Park: Software, Data curation, Visualization, Writing – original draft. Jaesung Choi: Investigation, Methodology, Software, Validation, Visualization, Writing – review and editing. Younghwa Cha: Data curation. UnCheol Lee: Methodology, Data curation. Joon-Young Moon: Conceptualization, Funding acquisition, Project administration, Supervision, Validation, Writing – review and editing. Jong-Min Park: Conceptualization, Funding acquisition, Project administration, Supervision, Validation, Writing – review and editing. All authors have read and agreed to the published version of the manuscript.
	
	\section*{Acknowledgements}
	\noindent ALC and JMP acknowledge research support from the JRG program at the APCTP, funded by the Science and Technology Promotion Fund and Lottery Fund of the Korean Government, with additional support from the local governments of Gyeongsangbuk-do Province and Pohang City. This work is also supported by the National Research Foundation (NRF) of Korea grant funded by the Korea government (MSIT) (RS-2025-00557038) (ALC and JMP). JC was supported by a KIAS Individual Grant (No. AP092902) via the Center for AI and Natural Sciences at the Korea Institute for Advanced Study (KIAS). This work is supported by the Center for Advanced Computation at KIAS. This research is also supported by IBS-R015-Y3, and the Basic Science Research Program through the National Research Foundation (NRF) of Korea, funded by the Ministry of Education (RS-2023-00272652). ALC was also partially supported by the National Research Foundation (NRF) of Korea under grant No. RS-2024-00343900.
	
	\section*{Conflict of Interest}
	\noindent The authors have no conflicts to disclose.
	
	\section*{Data Availability Statement}
	The data that support the findings of this study are available from the corresponding authors upon reasonable request.

    \section*{Declaration of generative AI and AI-assisted technologies in the manuscript preparation process}
	During the preparation of this work, the authors utilized ChatGPT (OpenAI) to assist with refining English expressions. After using this tool, the authors reviewed and edited the content as needed and take full responsibility for the content of the published article.
	
	\appendix

    \renewcommand{\thefigure}{\thesection.\arabic{figure}}
    \setcounter{figure}{0}
    
    \section{Details of Datasets}
        \label{sec:dods}
        In this section, we provide the details of the EEG datasets used in our study: (i) Dataset~\hyperref[sec:two_a]{I} comprising EEG recordings during general anesthesia obtained from the University of Michigan; and 
        (ii) Dataset~\hyperref[sec:two_b]{II} containing EEG data of individuals with inADHD from the Healthy Brain Network.

        \subsection{Dataset I: General Anesthesia}
        \label{sec:two_a}
        The dataset~\hyperref[sec:two_a]{I} consists of EEG recordings from eighteen healthy volunteers (aged 20-40 years) collected at the University of Michigan~\cite{kim2021criticality,lee2019relationship}. Among them, nine participants underwent general anesthesia administration, while the remaining were recorded without anesthesia. The study was reviewed in accordance with the recommendations of the Institutional Review Boards specializing in human subject research at the University of Michigan, Ann Arbor (Protocol \verb|#|HUM0071578).
	Written informed consent was obtained from all participants in accordance with the Declaration of Helsinki.
	In particular, we systematically extract seven distinct brain states during the general anesthesia protocol in the order in which they appear and investigate the degree of disorder in the brain dynamics in these states, capturing the different levels of consciousness: 
	\begin{itemize}
		\item Eyes Closed (EC): 5-minute eyes-closed resting state before anesthesia
		\item Propofol injection period (P): 5-minute post-propofol injection
		\item Loss of Consciousness (LOC): 3-minute post-LOC marker
		\item Burst period (B): 5-minute extracted burst period
		\item Suppression period (S): 5-minute extracted suppression period
		\item Post-Burst/Suppression anesthesia (PBS): 5-minute pre-recovery final state 
		\item Recovery of Consciousness (ROC): 5-minute post-ROC marker, return of consciousness
	\end{itemize}

        \subsection{Dataset II: Attention Deficit Hyperactivity Disorder (ADHD)}
	\label{sec:two_b}
	The Healthy Brain Network (HBN) dataset is released by the Child Mind Institute~\cite{alexander2017open}. ADHD is typically categorized into three main subtypes: inattentive, hyperactive-impulsive, and combined. To establish a more robust sample, we have selected individuals diagnosed with inattentive type ADHD (abbreviated as inADHD) without comorbidity. The control group consists of healthy individuals who have not received any diagnosis. Since the alpha spectrum peak can vary around the age of 10, we set the age range to 11 years and older~\cite{cragg2011maturation}. Briefly, we analyze resting-state EEG data from 40 participants with inADHD (34 males; mean age 13.97, s.e. 0.29) and 66 healthy control participants (34 males, mean age 14.06, s.e. 0.32). The resting-state protocol consists of five blocks, each comprising 20 seconds of eyes-open followed by 40 seconds of eyes-closed.

\section{EEG Data Acquisition and Preprocessing}
	\label{sec:edap}
	Both datasets~\hyperref[sec:two_a]{I} and ~\hyperref[sec:two_b]{II} were recorded using 128-channel HydroCel nets with Net Amps 400 amplifiers (Electrical Geodesic, Inc., USA) at a sampling rate of 500 Hz. We briefly describe our preprocessing procedure using the \texttt{EEGLAB v2022.1} package in MATLAB~\cite{delorme2004eeglab} as follows: (1) To remove 60 Hz due to the power line noise, we conduct notch filtering with the range 59-61 Hz using the \texttt{pop\_eegfiltnew.m} function. (2) To eliminate the global trend in a low-frequency band and artifact noise in a high-frequency band, we detrend with the range 0.5-100 Hz using the same \texttt{pop\_eegfiltnew.m} function. 
	(3) During the EEG recordings, some electrodes can have anomalous signals due to bad connections or participants' movements. We remove such bad channels using the \texttt{trimOutlier.m} function with the range 0.0001-100 $\mathrm{\mu V}$.
    (4) Furthermore, we extract the alpha band (8–12 Hz) using the \texttt{pop\_eegfiltnew.m} function. (5) Relative phase signals were calculated, and whole-brain topographic maps were generated by averaging the signals over 100 ms time windows. (6) Finally, regression analysis was applied to enhance computational efficiency and robustness.
    Please see reference~\cite{Park2025.03.12.642768} for more details.
    
	\section{Details of methods}
	\label{app:A3}
\subsection{Time series from relative phase dynamics of EEG signals}
	\label{sec:two_d}
When relative phase analysis is applied to the EEG time series data from a 128-channel system, it yields a relative phase map at each time point, indicating which regions of the brain are phase-leading and which are phase-lagging (Fig.~\ref{fig:workflow_pipeline}(b)). Extracting meaningful phase information from EEG electrodes presents two primary challenges. First, phase discontinuity arises due to the periodic nature of phase values, as the phase of a signal, $\theta$, or the phase difference between signals, $\Delta \theta$, is defined in the range of (-$\pi$, $\pi$) where -$\pi$ and $\pi$ are equivalent. Second, due to potential artifacts from volume conduction and other noise sources in EEG signals, small phase differences between electrodes, especially when $\Delta\theta \sim 0$, may not reflect true neural interactions. To address both issues, we apply a sine transformation to the phase differences $\theta_j(t) - \Omega(t)$. This approach assigns the greatest weight to phase differences of $\pm \pi/2$ while minimizing the influence of values near $\pi$ and $0$, thereby enhancing the robustness of the measure against values near $\pi$ and $0$~\cite{vinck2011improved}. 

 From our previous study~\cite{Park2025.03.12.642768}, we have identified four dominant patterns that account for a substantial fraction of the variance in these relative phase maps: (i) the anterior regions leading the posterior; (ii) the left hemisphere leading the right; (iii) the right hemisphere leading the left; and (iv) the posterior regions leading the anterior. These dominant modes can be robustly extracted using either K-means clustering or principal component analysis (PCA) (see Fig.~1 A-C and Fig.~5 A-B of Ref.~\cite{Park2025.03.12.642768}). When PCA is applied to the relative phase time series, the anterior-posterior directionality (patterns (i) and (iv)) and the left-right directionality (patterns (ii) and (iii)) emerge as the first and second principal components, respectively. Notably, the first principal component accounts for approximately 50\% of the total variance, while the second explains about 25\%~\cite{Park2025.03.12.642768}.

 \subsection{Meaning of $\beta_1$}
        \label{app:mb}
        We explain the meanings of the weights $\beta_1$ and $\beta_2$ along the first $\boldsymbol{X}_1$ and second $\boldsymbol{X}_2$ principal axes in Eq.~\eqref{eq:sb}. While $\boldsymbol{X}_1$ represents the anterior-posterior directionality regressor, $\boldsymbol{X}_2$ represents the left-right directionality regressor. This regression yields two global time series, $\beta_1(t)$ and $\beta_2(t)$, capturing the temporal fluctuations in the expression of the anterior-posterior and left-right phase patterns, respectively (Fig.~\ref{fig:workflow_pipeline}(c)).

The sign of $\beta_1(t)$ provides directional information: positive values indicate anterior-to-posterior dominance, while negative values indicate posterior-to-anterior dominance. Similarly, the sign of $\beta_2(t)$ reflects lateralization, with positive values corresponding to left-to-right dominance and negative values to right-to-left dominance. Thus, $\beta_1(t)$ and $\beta_2(t)$ offer concise, continuous measures of large-scale directional phase dynamics in the brain.

        \subsection{Permutation entropy}
        \label{sec:pee}
We provide a detailed explanation of the steps used in computing ordinal patterns and permutation entropy (Fig.~\ref{fig:workflow_pipeline}(d, e)), accompanied by an illustrative example using embedding dimension $dx = 3$ as follows.

\begin{figure}
\includegraphics[width=0.48\textwidth]{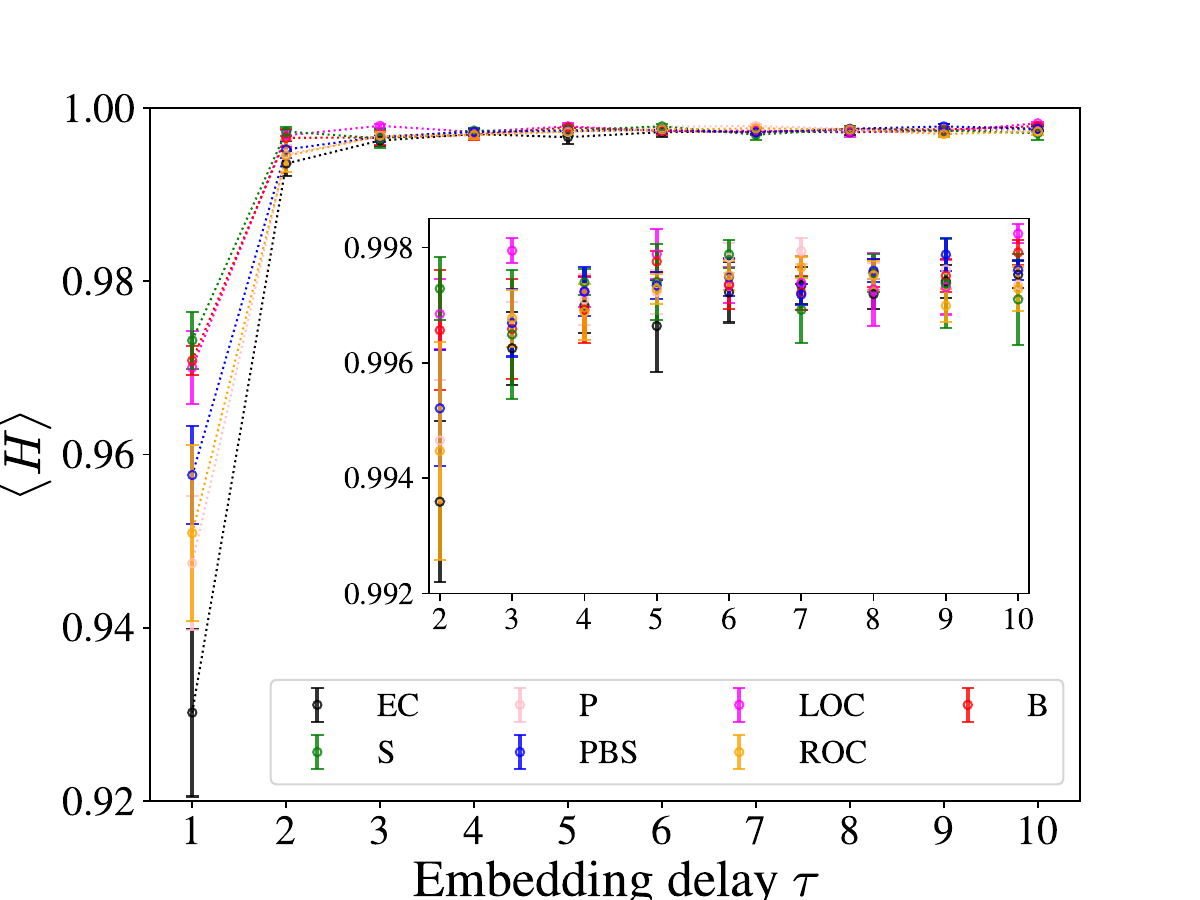}
\vskip -0.1in
\caption{
\textbf{Embedding delay $\tau$ dependence of mean $\beta_1$-PE $\langle H \rangle$ obtained from the general anesthesia dataset~\hyperref[sec:two_a]{I}:} Symbols represent the mean $\beta_1$-PE $\langle H \rangle$ for time segments of length $1500$ with $dx =4$. Brain states are represented by the same colors as in Fig.~\ref{fig:beta1_pe_time_length}. The inset provides a magnified view of the results for $\tau > 1$.
}
		\label{fig:tau_dependence}
\end{figure}

\FloatBarrier

\setcounter{figure}{0}
\renewcommand{\thefigure}{D.\arabic{figure}}

\begin{figure*}
\includegraphics[width=\textwidth]
{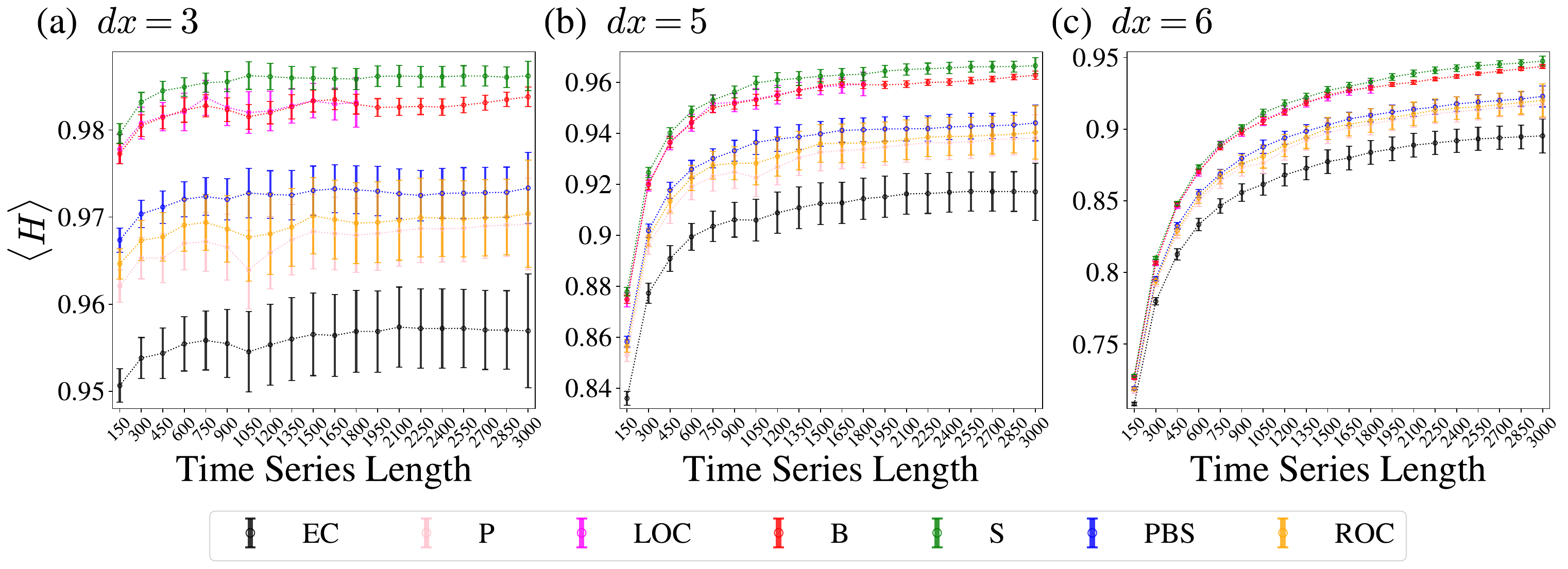}
\vskip -0.1in
\caption{\textbf{Time-series length dependence of $\beta_1$-PE across different brain states in the general anesthesia dataset~\hyperref[sec:two_a]{I}:} Symbols represent the mean $\beta_1$-PE $\langle H \rangle$ calculated using (a) $d x = 3$, (b) $dx = 5$, and (c) $dx = 6$. Brain states are represented by the same colors as in Fig.~\ref{fig:beta1_pe_time_length}.} 
		\label{fig:dx_variation_time_length}
	\end{figure*}

\begin{figure*}
\includegraphics[width=\textwidth]
{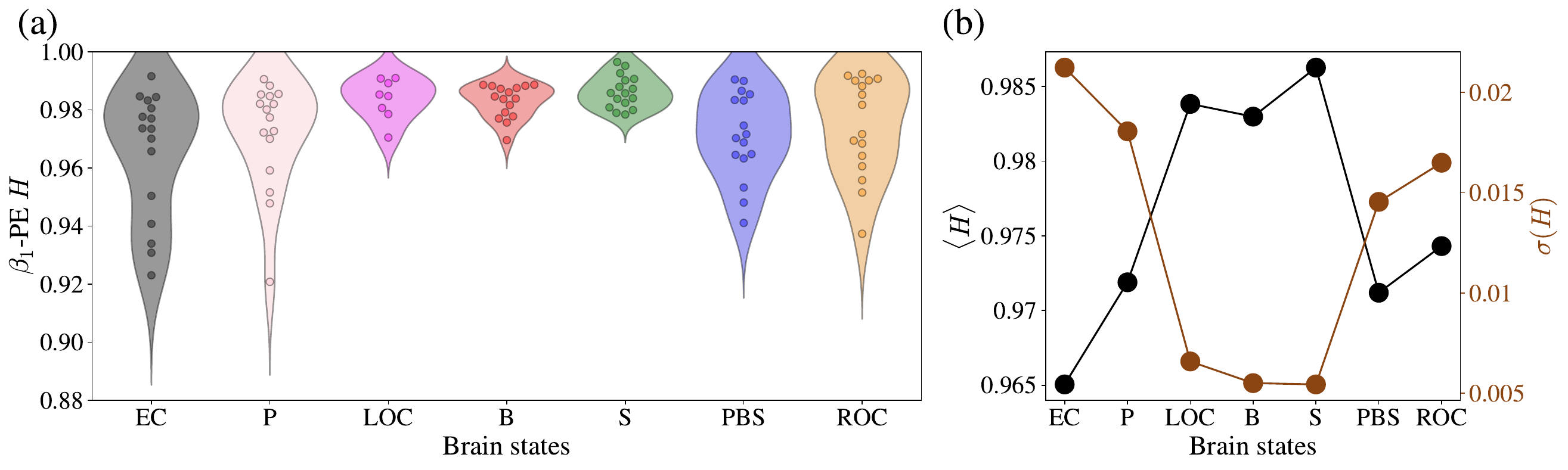} 
\vskip -0.1in
\caption{\textbf{Statistical properties of $\beta_1$-PE for individuals exhibiting all brain states in the general anesthesia dataset~\hyperref[sec:two_a]{I}:}
(a) Violin plots and symbols represent the distribution of $\beta_1$-PE for time segments of length 750 with $d x = 3$.
(b) Symbols indicate the mean and standard deviation of $\beta_1$-PE computed over time segments. Brain states and statistical measures are represented by the same colors as in Fig.~\ref{fig:beta1_pe_distribution}.}
		\label{fig:distribution_subset}
	\end{figure*}

The Bandt-Pompe procedure~\cite{bandt2002permutation,pessa2021ordpy,chanu2024exploring,chanu2025climate} for computing ordinal patterns and their distribution follows these steps:
	\begin{enumerate}
		\item For a time series $\mathcal{X}=\{x_i~;~i=1,2,3, \dots,M\}$ of given length $M$, we divide it into overlapping partitions $m=M-(dx-1)\tau$ with embedding delay $\tau$. Our analysis takes consecutive time units ($\tau=1$). 
		\item Next, for each data partition $\mathcal{D}_p=(x_p,x_{p+1},\dots,x_{p+(dx-1)})$ with partition index $p=1,2,3,\dots,m$, we determine a permutation state $\pi_p=(u_0,u_1,\dots,u_{dx-1})$ by sorting the elements in ascending order. Specifically, the inequality $x_{p+u_0}\leq x_{p+u_1}\leq \dots \leq x_{p+u_{dx-1}}$ defines the permutation of the index numbers.
		\item Now, we generate the symbolic sequence $\{\pi_p\}_{p=1,2,3,\dots,m}$ known as ordinal pattern sequence.
		\item We calculate the relative frequency of all possible patterns as:
		\begin{equation}
			\rho(\pi_j)= \frac{\textrm{\# patterns of type } \pi_j}{m},
		\end{equation}
		where $P=\{\rho(\pi_j)\}$ is the ordinal probability distribution with $j=1,2,3,\dots,dx!$. 
	\item Then we compute the permutation entropy $S[P]$~\cite{bandt2002permutation} and the normalized permutation entropy $H[P]$~\cite{rosso2007distinguishing}.
	\end{enumerate}
For computing $H$, we employ the open-source Python module \texttt{ordpy}~\cite{pessa2021ordpy}.

Previous studies have shown that larger embedding delays can provide additional insight into the underlying dynamics~\cite{parlitz2012classifying, zunino2012distinguishing}.
In our analysis, we set $\tau = 1$, as the EEG data are already downsampled to a relatively low sampling frequency.
For the present EEG data, the effective sampling frequency becomes too low to reliably extract meaningful features of the brain states when $\tau \ge 2$. As a result, the permutation entropy rapidly saturates near its maximum value across all brain states, thereby reducing its discriminative power, as shown in Fig.~\ref{fig:tau_dependence}.

\setcounter{figure}{0}
\renewcommand{\thefigure}{E.\arabic{figure}}

\begin{figure*}
\includegraphics[width=\textwidth]{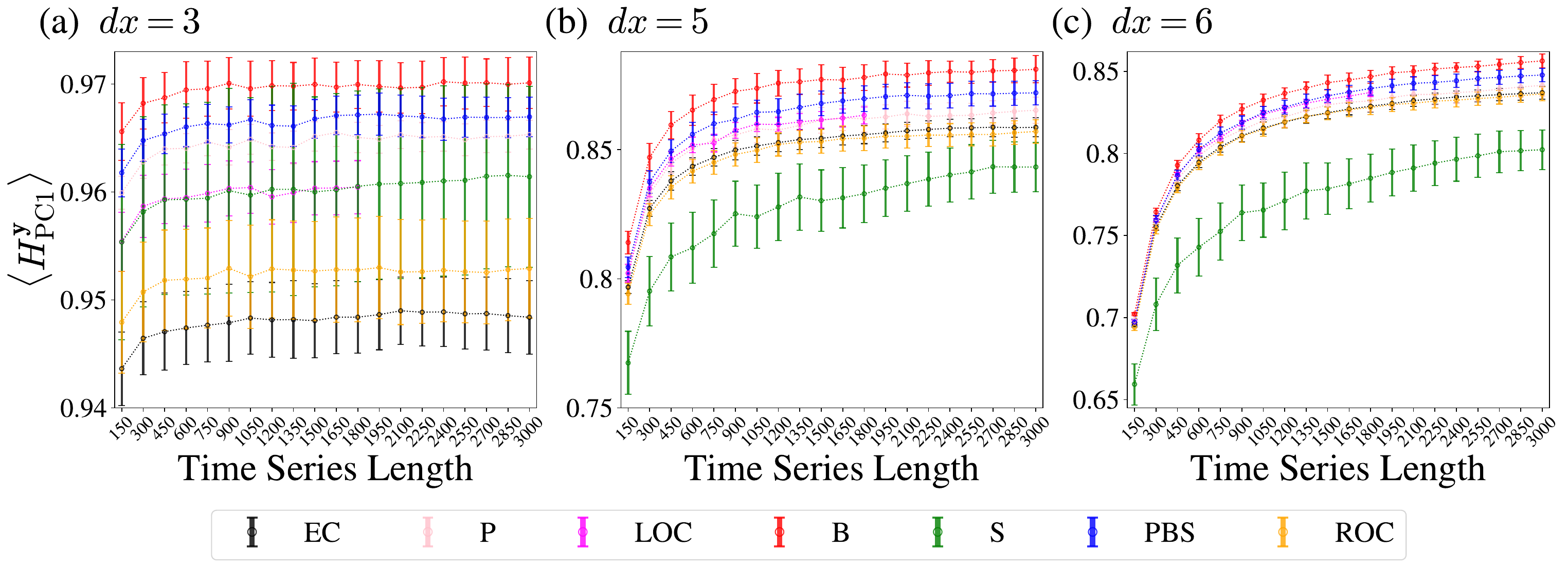}
\vskip -0.1in
\caption{
\textbf{Time-series length dependence of mean raw-data-based permutation entropy across different brain states in the general anesthesia dataset~\hyperref[sec:two_a]{I}:}
Symbols represent the mean permutation entropy $\langle H^{\boldsymbol{y}}_{\mathrm{PC}1} \rangle$ calculated using (a) $d x = 3$, (b) $dx = 5$, and (c) $dx = 6$. Brain states are represented by the same colors as in Fig.~\ref{fig:beta1_pe_time_length}.
}
		\label{fig:raw_dx_variation_time_length}
	\end{figure*}

	To illustrate, consider the time series $\mathcal{X}=(44,18,10,7,32,14)$ with $dx=3$ and $\tau=1$. This generates $dx!=3!=6$ possible $\{\pi_j\}$ permutations: $\pi_1=(0,1,2), \pi_2=(0,2,1), \pi_3=(1,0,2), \pi_4=(1,2,0), \pi_5=(2,0,1)$, and $\pi_6=(2,1,0)$. We analyze each partition as follows: 
	\begin{itemize}
		\item Sorting the elements of $\mathcal{D}_1=(44,18,10)$ in ascending order yields $10<18<44$, indicating $x_{p+2}<x_{p+1}<x_{p}$. Hence, the ordinal pattern associated with $\mathcal{D}_1$ is $\pi_6=(2,1,0)$. 
		\item Likewise, for $\mathcal{D}_2=(18,10,7)$, sorting yields $7<10<18 $, corresponding to $\pi_6=(2,1,0)$.
		\item Continuing for $\mathcal{D}_3$ and $\mathcal{D}_4$ results in the final symbolic sequence $\{\pi_p = \pi_6,\pi_6,\pi_3,\pi_2\}$. 
        \item By counting the frequency of each pattern, we obtain
\begin{equation}
    \rho(\pi_j) = \left \{
    \begin{matrix*}[l]
    \frac{1}{2}, & \textrm{for } j = 6, \\
    \frac{1}{4}, & \textrm{for } j = 2,3, \\
    0, & \textrm{otherwise}. \\
    \end{matrix*}
    \right .
\end{equation}
		\item  
        Finally, we get $S=1.5$ and $H=0.5802$.
	\end{itemize}

     \section{$\beta_1$-PE calculated using different embedding dimension $dx$ in dataset~\hyperref[sec:two_a]{I}}
        \label{sec:extra}

\begin{figure}
\includegraphics[width=0.45\textwidth]{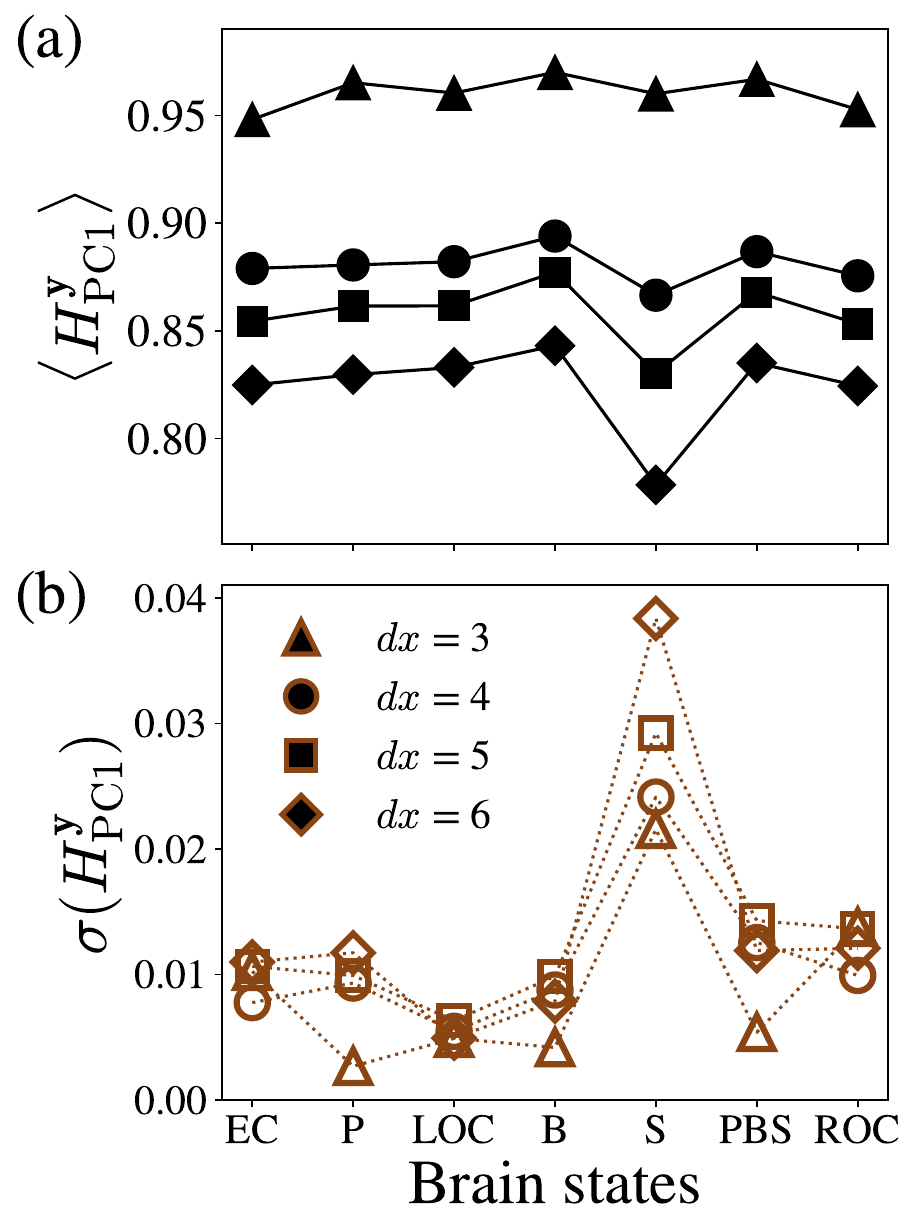}
\vskip -0.1in
\caption{
\textbf{Mean and standard deviation of the raw-data-based permutation entropy distribution in the general anesthesia dataset~\hyperref[sec:two_a]{I}:}
Black and brown symbols indicate (a) the mean and (b) standard deviation of the raw-data-based permutation entropy computed over the time segments, respectively. The symbol shapes correspond to $dx = 3$ (triangle), $dx = 4$ (circle), $dx = 5$ (square), and $dx = 6$ (diamond).
}
		\label{fig:raw_dx_statistics}
	\end{figure}

\begin{figure*}
\includegraphics[width=\textwidth]{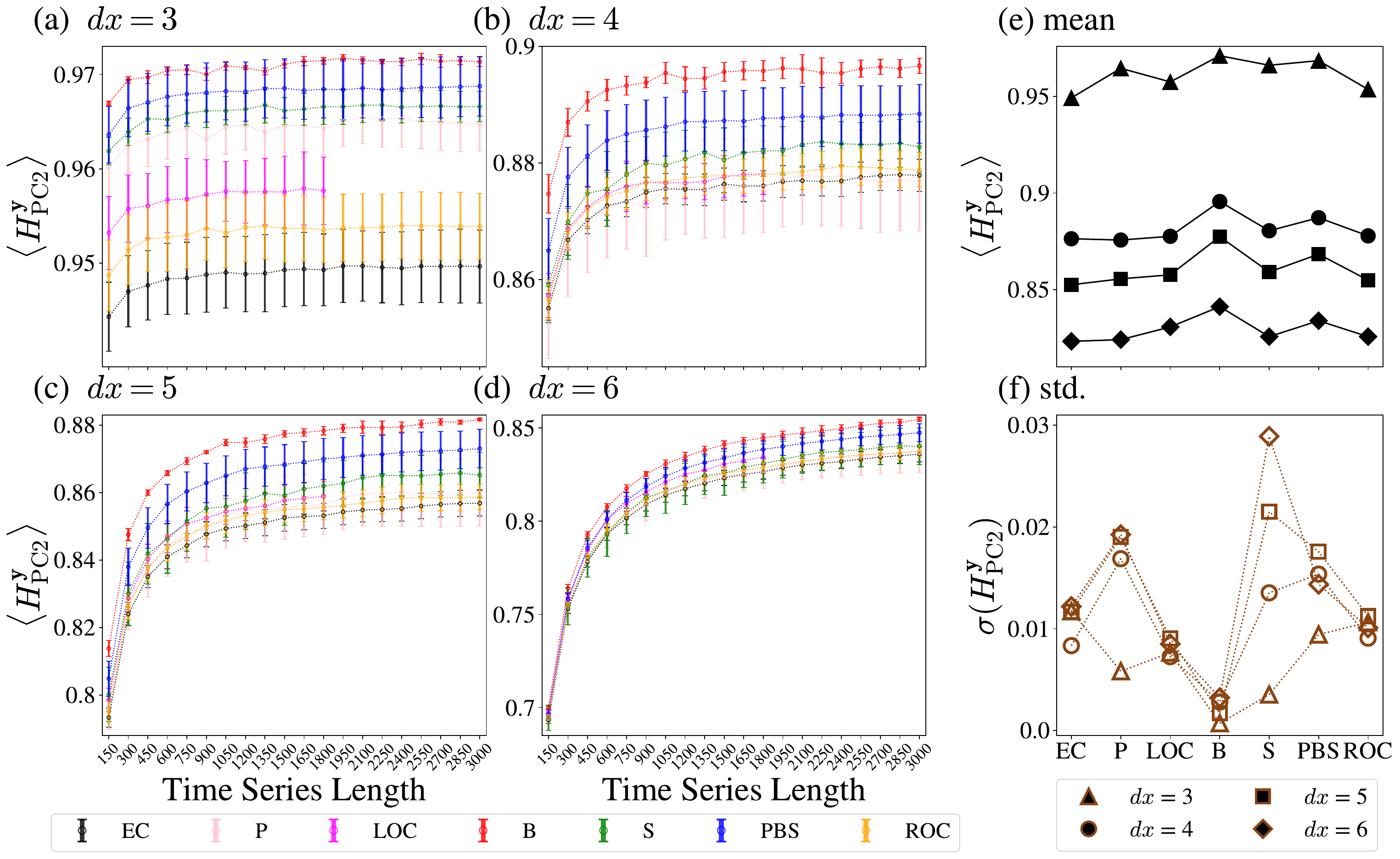}
\vskip -0.1in
\caption{
\textbf{Results for raw-data-based permutation entropy computed from the second principal component in the general anesthesia dataset~\hyperref[sec:two_a]{I}:}
(a-d) Symbols represent the mean permutation entropy $H^{\boldsymbol{y}}_{\mathrm{PC}2}$ computed from the second principal component in the raw EEG data using (a) $d x = 3$, (b) $d x = 4$, (c) $d x = 5$, and (d) $d x = 6$. Brain states are represented by the same colors as in Fig.~\ref{fig:beta1_pe_time_length}.
(e-f)
Black and brown symbols indicate (e) the mean and (f) standard deviation of the raw-data-based permutation entropy computed from the second principal component over the time segments, respectively. The symbol shapes correspond to $dx = 3$ (triangle), $dx = 4$ (circle), $dx = 5$ (square), and $dx = 6$ (diamond).
}
		\label{fig:raw_pe_pca2}
	\end{figure*}

To validate the robustness of our results for dataset~\hyperref[sec:two_a]{I} against different choices of the embedding dimension $d x$, we investigate the time-series length dependence of $\beta_1$-PE for $d x = 3$, $5$, and $6$, as shown in
Fig.~\ref{fig:dx_variation_time_length}.
The results indicate that the overall qualitative trends remain nearly identical across different values of $d x$, except for quantitative differences in the saturation point and the error magnitude. For smaller values of $d x$, saturation occurs relatively rapidly, allowing values comparable to those obtained from long-duration recordings to be extracted even from shorter time series.
In contrast, for larger $d x$, saturation occurs much more slowly, requiring longer recordings to reach values independent of the time-series length.
However, larger values of $d x$ yield smoother curves with smaller errors.
These results demonstrate that the choice of a proper embedding dimension $d x$ may carefully balance the trade-off between saturation time and statistical reliability.

To further verify that the statistical properties of $\beta_1$-PE are not significantly dependent on the embedding dimension $d x$, we examine $\beta_1$-PE using $d x = 3$ and time segments of length 750.
We further investigate the influence of individual subjects who did not reach deep unconscious states and therefore did not exhibit all brain states.
To do this, we repeat the analysis using only the five subjects who exhibited all brain states.
The resulting violin plots and associated statistics are shown in Fig.~\ref{fig:distribution_subset}. We observe behavior consistent with that noted in Fig.~\ref{fig:beta1_pe_distribution}.
These observations demonstrate
that our findings are robust with respect to both the choice of embedding dimension and the potential absence of deep unconscious states in subjects.

\setcounter{figure}{0}
\renewcommand{\thefigure}{F.\arabic{figure}}

\begin{figure*}
\includegraphics[width=\textwidth]{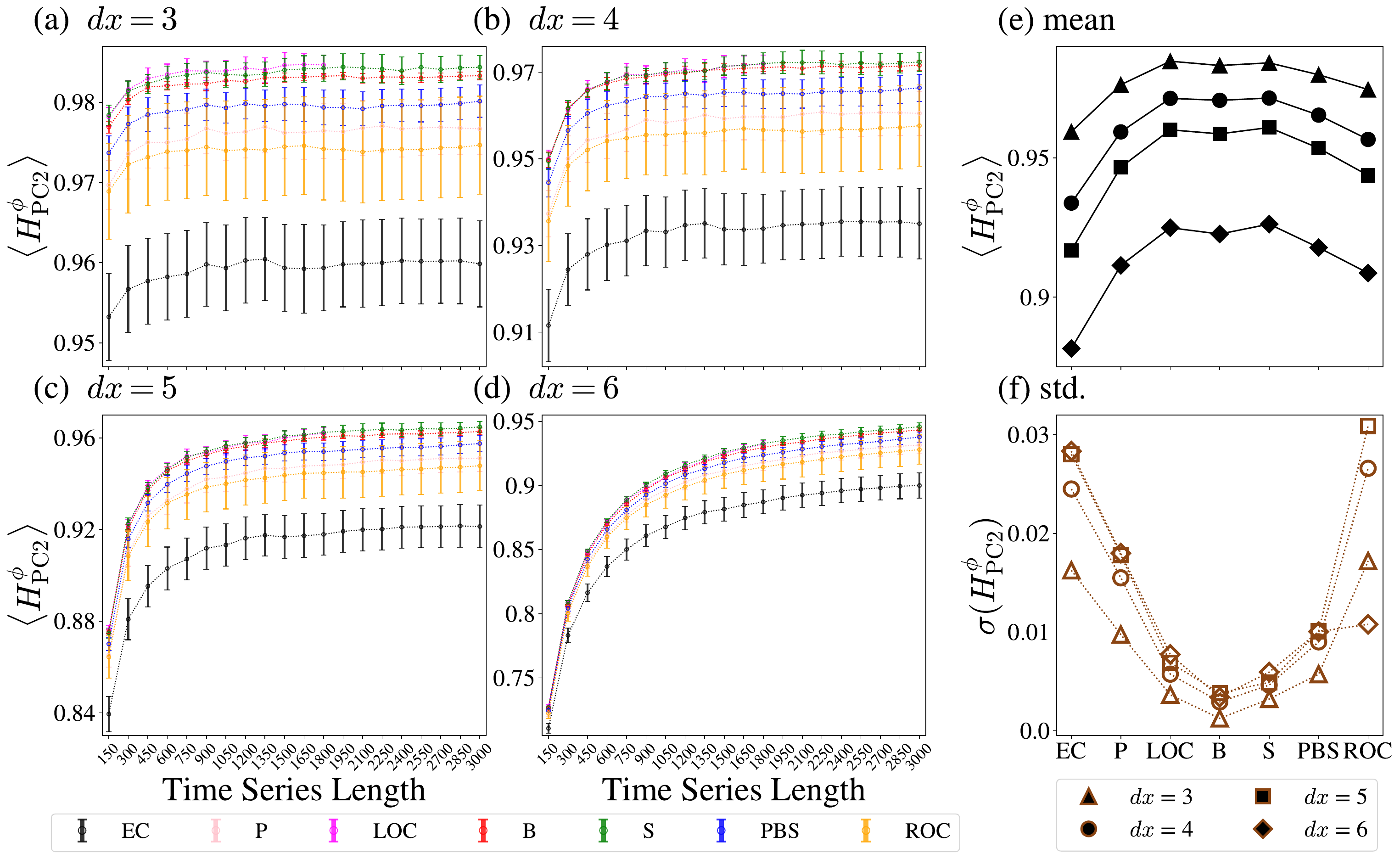}
\vskip -0.1in
\caption{
\textbf{Results for $\beta_2$-PE in the general anesthesia dataset~\hyperref[sec:two_a]{I}:}
(a-d) Symbols represent the mean $\beta_2$-PE $\langle H^{\phi}_{\mathrm{PC}2} \rangle$ calculated using (a) $d x = 3$, (b) $d x = 4$, (c) $d x = 5$, and (d) $d x = 6$. Brain states are represented by the same colors as in Fig.~\ref{fig:beta1_pe_time_length}.
(e-f)
Black and brown symbols indicate (e) the mean and (f) standard deviation of the $\beta_2$-PE over the time segments, respectively. The symbol shapes correspond to $dx = 3$ (triangle), $dx = 4$ (circle), $dx = 5$ (square), and $dx = 6$ (diamond).
}
		\label{fig:beta2_time_length}
	\end{figure*}

\section{Limitations of Raw-Data-Based Permutation Entropy}
        \label{sec:review}

In this Appendix, we discuss the limitations that arise when the same analysis is applied directly to the raw EEG signals without transforming the data into phase dynamics. In particular, we demonstrate that permutation entropy computed from the raw data does not exhibit a systematic correlation with brain states.

First, we investigate the time-series length dependence of the raw-data based permutation entropy $H^{\boldsymbol{y}}_{\mathrm{PC}1}$ for several values of the embedding dimension $d x$. Fig.~\ref{fig:raw_dx_variation_time_length} shows the results for $dx = 3, 5$ and $6$. As noted in the main text, the permutation entropy computed from the raw data converges more slowly compared to $\beta_1$-PE, indicating that a larger number of data points is required to reliably estimate the ordinal pattern statistics.

In addition, the raw-data-based permutation entropy exhibits a strong dependence on the embedding dimension $d x$. In particular, the relative ordering of permutation entropy across brain states is not preserved when $d x$ is varied, indicating that the observed behavior is sensitive to the choice of embedding parameter.

To further illustrate this point, Fig.~\ref{fig:raw_dx_statistics} presents the mean and standard deviation of the raw-data-based permutation entropy across brain states, evaluated at a fixed time-series length of 1500 for different values of $d x$. 
The mean and standard deviation exhibit several undesirable properties: (i) a strong dependence on $d x$, (ii) no correlation with the level of consciousness, and (iii) the absence of the previously observed anti-correlation between them.

As an additional check, we perform the same analysis using the second principal component. Figure~\ref{fig:raw_pe_pca2} shows that it exhibits similar behavior, indicating that these observations are not specific to the leading principal component.

Overall, these results indicate that permutation entropy computed directly from raw EEG data is highly sensitive to the choice of embedding parameter and does not provide a consistent characterization of brain states. In contrast, the phase-based representation yields a low-dimensional signal that preserves state-dependent structure and enables robust estimation of ordinal pattern statistics.

\section{Analysis of $\beta_2(t)$}
        \label{sec:beta2}

\begin{figure}
\includegraphics[width=0.45\textwidth]{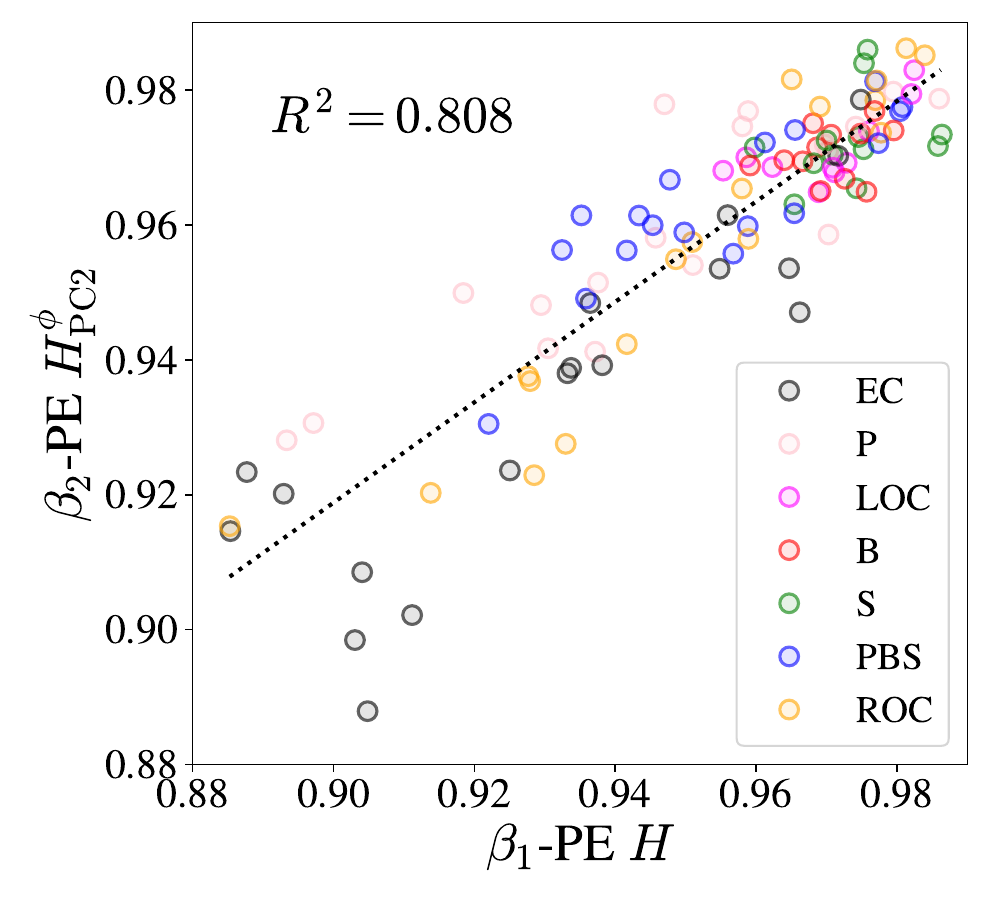}
\vskip -0.1in
\caption{
\textbf{Correlation between $\beta_1$-PE and $\beta_2$-PE in the general anesthesia dataset~\hyperref[sec:two_a]{I}:}
Both $\beta_1$-PE and the $\beta_2$-PE are computed for the same time segment length $1500$ and $dx=4$. Brain states are represented by the same colors as in Fig.~\ref{fig:beta1_pe_time_length}.
The black dashed line indicates the result of a linear fit, yielding the coefficient of determination $R^2 = 0.808$.
}
		\label{fig:beta2_distribution}
	\end{figure}

We now present the results of the same analysis performed on the second principal mode, $\beta_2(t)$, in order to assess whether it provides additional information beyond that obtained from $\beta_1(t)$.

The results are summarized in Fig.~\ref{fig:beta2_time_length}.
In particular, Fig.~\ref{fig:beta2_time_length}(a)-(d) show that the permutation entropy of $\beta_2(t)$, denoted by $\beta_2$-PE, exhibits behavior similar to that of $\beta_1$-PE.
The convergence behavior is comparable with a similar saturation time series length and the relative ordering is robust with respect to $dx$.
Overall, the behavior remains qualitatively similar, with only minor differences, such as slightly larger error bars and small variations in the ordering between P and ROC.

In addition, Fig.~\ref{fig:beta2_time_length}(e) and (f) show that the mean and standard deviation of $\beta_2$-PE exhibit both clear correlations with the level of consciousness and an anti-correlation between them.

We further find that $\beta_1$-PE and $\beta_2$-PE are strongly correlated, as shown in Fig.~\ref{fig:beta2_distribution}. As a result, incorporating $\beta_2$-PE does not provide additional independent information beyond that already captured by $\beta_1$-PE.

These results indicate that, while $\beta_2$-PE exhibits qualitatively similar and robust behavior, it does not contribute new information for characterizing brain states. Therefore, focusing on either $\beta_1(t)$ or $\beta_2(t)$ alone is sufficient for the analysis.

    \FloatBarrier
    \bibliographystyle{refstyl}
	\bibliography{main}

\end{document}